\pdfoutput=1

\documentclass[11pt,twoside,a4paper,cmspaper,final,collab]{cms-tdr}
\def\svnVersion{175808}\def\cmsCernNo{2012-297}\def\cmsCernDate{\today}\def\cmsMessage{Submitted to the Journal of High Energy Physics}

\begin{document}\cmsNoteHeader{EXO-11-030}

\hyphenation{had-ron-i-za-tion}
\hyphenation{cal-or-i-me-ter}
\hyphenation{de-vices}
\RCS$HeadURL: svn+ssh://svn.cern.ch/reps/tdr2/papers/EXO-11-030/trunk/EXO-11-030.tex $
\RCS$Id: EXO-11-030.tex 154160 2012-10-23 01:32:42Z alverson $
\providecommand\sbottom{\ensuremath{\widetilde{\cmsSymbolFace{b}}_1}\xspace}
\providecommand\neutralino{\PSGczDo\xspace}
\providecommand\etmiss{\ETslash}
\providecommand{\re}{\ensuremath{\cmsSymbolFace{e}}\xspace}

\cmsNoteHeader{EXO-11-030}

\title{Search for third-generation leptoquarks and scalar bottom quarks in pp collisions at $\sqrt{s}= 7\TeV$}

\date{\today}

\abstract{
  Results are presented from a search for third-generation leptoquarks and scalar
  bottom quarks in a sample of proton-proton collisions at $\sqrt{s} = 7\TeV$ collected by the CMS experiment at the LHC, corresponding to an
  integrated luminosity of 4.7\fbinv. A scenario where the new particles are pair produced and each decays to a b quark plus a tau neutrino or neutralino is considered. The number of  observed events is found to be in agreement with the standard model prediction. Upper limits are set at 95\% confidence level on the production cross sections. Leptoquarks with masses below ${\sim}450$\GeV are excluded.  Upper limits in the mass plane of the scalar quark and neutralino  are set such that scalar bottom quark masses up to 410\GeV are excluded for neutralino masses of 50\GeV.
  }

\hypersetup{%
  pdfauthor={CMS Collaboration},%
  pdftitle={Search for third-generation leptoquarks and scalar bottom quarks in pp collisions at sqrt(s) = 7 TeV},%
  pdfsubject={CMS},%
  pdfkeywords={CMS, MET, razor, leptoquark}}

\maketitle

\section{Introduction}
 \label{sec:intro}
 Many theoretical extensions of the standard model (SM) predict the existence of color-triplet scalar or vector bosons, called leptoquarks (LQ), that have fractional electric charge and both lepton and baryon quantum numbers. These theories include grand unified theories~\cite{PatiSalam}, composite models~\cite{Schrempp1985101, Gripaios:2010hv}, technicolor
schemes~\cite{Dimopoulos1979237, Dimopoulos198069,Farhi1981277}, and
superstring-inspired $E_6$ models~\cite{Hewett1989193}.  We follow the usual assumption that there are
three generations of LQs, each of which couples only to the
corresponding generation of SM particles, to avoid violating the known
experimental constraints on flavor-changing neutral currents~\cite{DavidsonBaileyCampbell}. Leptoquarks would be produced at the Large Hadron Collider (LHC) in pairs predominantly through $\Pg\Pg$ fusion and $\cPq\cPaq$ annihilation, and the contributions from lepton $t$-channel exchange are suppressed by the leptoquark Yukawa couplings. A leptoquark decays to a charged lepton and a quark with a branching fraction $\beta$ usually considered as a free parameter of the model, or a neutrino and a quark with branching fraction $1-\beta$. For scalar LQs, the production cross section
is determined by the ordinary color coupling between an LQ and a gluon, which is
model independent.

Numerous theories of particle physics beyond the SM address the gauge hierarchy problem and other shortcomings of the SM by
introducing a new symmetry that relates fermions and bosons, called ``supersymmetry'' (SUSY)~\cite{Farrar1978575}. Supersymmetric models introduce a new discrete symmetry, R-parity, and all SM particles have $R_p=+1$ while all superpartners have $R_p =-1$. Imposing R-parity conservation prohibits baryon and lepton number violating couplings which could otherwise lead to rapid  proton decay. In models with
R-parity conservation, SUSY particles are produced in pairs, and the
lightest SUSY particle (LSP) is stable. In some models the LSP is the electrically neutral and weakly interacting neutralino (\neutralino), which provides a dark
matter candidate~\cite{Feng:2010gw}.  The left- and right-handed SM
quarks have scalar partners ($\tilde{\cPq}_L$ and $\tilde{\cPq}_R$) that can
mix to form scalar quarks (squarks) with mass eigenstates
$\tilde{\cPq}_{1,2}$. Since the mixing is proportional to the corresponding SM
fermion masses, the effects can be enhanced for the third generation
squarks, yielding sbottom ($\tilde{\cPqb}_{1,2}$) and stop ($\tilde{\cPqt}_{1,2}$) mass
eigenstates with large mass splitting. The lighter mass eigenstate ($\tilde{\cPqb}_1$ or $\tilde{\cPqt}_1$) could be lighter than any other charged SUSY particle~\cite{Dimopoulos:1995mi}. Therefore, if sufficiently light, \sbottom squarks could be
 produced at the LHC either directly or through decays of
gluinos (the supersymmetric partners of gluons). In most SUSY models, a \sbottom is expected to decay predominantly into a bottom
quark and \neutralino, so that the final state consists of b jets and a sizable imbalance in transverse energy ($\etmiss$), defined as the magnitude of the vector opposite to the sum of the transverse momenta of all detected particles.

In this paper we present results of a search for pair-produced scalar
third-generation leptoquarks (LQ$_3$) with an electric charge of $\pm$1/3 and for $\tilde{\cPqb}_1$. Each of the LQ$_3$ ($\tilde{\cPqb}_1$) particles decays into a
b quark and $\nu_\tau$ (\neutralino). In each case, signal events
are characterized by two high-transverse-momentum ($\pt$) b jets accompanied by large $\etmiss$.  The resulting final state, consisting of jets, $\etmiss$, and no charged leptons, does not allow a full reconstruction
of the decay chain, because of the lack of knowledge of the individual momenta of the weakly interacting particles.

Previous searches performed by the CDF and D0 collaborations at the Tevatron have excluded
LQ$_3\to\nu_{\tau}\bar{b}$ masses below 247\GeV, and set limits on the production of \sbottom squarks
for a range of values in the $ \widetilde{\cPqb}_1- \neutralino$ mass plane that extend up to $m(\sbottom) = 200\GeV$
for $m(\neutralino) = 110\GeV$~\cite{Abazov201095,PhysRevLett.105.081802}. A search performed by the CMS
collaboration has excluded the existence of a scalar LQ$_3$ with an electric charge of $\pm$2/3 or $\pm$4/3 and with mass below 525\GeV, assuming 100\% branching fraction to a b quark and a $\tau$ lepton~\cite{CMS-PAS-EXO-12-002}. A search performed by the ATLAS collaboration excluded the production of \sbottom with masses up to 390\GeV, for \neutralino masses below 60\GeV~\cite{PhysRevLett.108.181802}.

The main SM backgrounds in this search are $\ttbar$+jets, heavy-flavor
(HF) multijet production, and $\PW$ or $\cPZ$ accompanied by HF
production. In the case of multijet events and $\PW/\cPZ$ decays to hadrons,
the $\etmiss$ is due to neutrinos in HF semileptonic decays, and due to effects of jet energy resolution and mismeasurements. In the case of $\PW/\cPZ$ decays to leptons, genuine $\etmiss$
results from the escaping neutrinos when the charged lepton ($\Pe$ or
$\mu$) goes undetected, or from $\tau$ decays.

\section{The CMS apparatus}
A detailed description of the Compact Muon Solenoid (CMS) detector can be found
elsewhere~\cite{PTDR1}. The central feature of the CMS detector is
the superconducting solenoid magnet, of 6\unit{m} internal diameter, providing
a magnetic field of 3.8\unit{T}. The silicon pixel and strip tracker, the
lead-tungstate crystal electromagnetic calorimeter (ECAL), and the brass/scintillator
hadron calorimeter (HCAL) are contained within the solenoid. Muons are
detected in gas-ionization chambers embedded in the steel return
yoke. The ECAL has a typical energy resolution of 1--2\% for electrons
and photons above 100\GeV. The HCAL, combined with the ECAL, measures the
jet energy with a resolution $\Delta E/E\approx100\mbox{\%}/\sqrt{E/\GeVns}
\oplus 5\mbox{\%}$.

CMS uses a right-handed coordinate system, with the origin located at
the nominal collision point, the $x$ axis pointing towards the center of
the LHC ring, the $y$ axis pointing up (perpendicular to the plane of LHC ring), and
the $z$ axis along the counterclockwise-beam direction. The azimuthal angle $\phi$ is
measured with respect to the $x$ axis in the $x$-$y$ plane and the polar
angle $\theta$ is defined with respect to the $z$ axis. The
pseudorapidity is defined as $\eta = -\ln [\tan(\theta / 2) ]$.

\section{Razor variables}
Although the signal considered in
this analysis consists of two high $\pt$ b jets and $\etmiss$, additional
jets may be produced by
initial- or final-state radiation (ISR/FSR). We study the
effect of such radiation with Monte Carlo (MC) simulation samples.  To reduce the
systematic uncertainty due to the imperfect simulation of ISR/FSR, we force every event into a
  dijet topology by combining all the jets in the event into two ``pseudojets'', following the ``razor'' methodology and variables~\cite{rogan, SUS-10-009}. The pseudojets are constructed as a sum of the four-momenta
of their constituent jets. After considering all
possible partitions of the jets into two pseudojets, the combination that minimizes the sum in quadrature of the pseudojet masses is selected.

The razor methodology provides an inclusive technique to search for production of heavy particles, each decaying to a visible system of particles and a weakly interacting particle. As an example, let us consider the pair production of two massive particles, denoted $S$, each decaying to a b quark and neutral weakly interacting particle, $\chi$, as $S \to \cPqb \chi$. In the respective rest frame of each particle $S$, the decay products have a unique momentum $p$ resulting from the two-body decay of $S$, given by:

\begin{equation}
p = \frac{M_S^2-M_\chi^2}{2M_S},
\end{equation}

where the mass of the b quark is neglected in this expression. This characteristic momentum, which is  denoted $M_{\Delta}$ and is referred to as ``momentum scale'', is the same in each decay instance, and can be used to distinguish this particular signal from SM backgrounds in the same final states. The razor mass, $M_{R}$, is an event-by-event estimator of this scale calculated through a series of approximations, motivated by physics, meant to estimate the rest frames of the respective particles $S$~\cite{rogan, SUS-10-009}, and is defined as:

\begin{equation}
M_{R} \equiv \sqrt{(|\vec{p^1}| +|\vec{p^2}|)^{2} - (p_{z}^{1}+p_{z}^{2})^{2}} \sim 2M_{\Delta},
\end{equation}

where $p^i$ ($p_z^i$) is the absolute value (the longitudinal component) of the $i$-th pseudojet momentum. An average transverse mass $M_\mathrm{T}^{R}$ can be defined as:
\begin{eqnarray}
  M_\mathrm{T}^{R}\equiv \sqrt{ \frac{\etmiss(p_\mathrm{T}^{1}+p_\mathrm{T}^{2}) - \vec{\etmiss} {\mathbf \cdot}(\vec{p}_\mathrm{T}^{1}+\vec{p}_\mathrm{T}^{2})}{2}} ,
\end{eqnarray}
whose maximum value for signal events equals $M_{\Delta}$. The dimensionless
variable $R$ is then defined as:
\begin{eqnarray}
  R \equiv \frac{M_\mathrm{T}^{R}}{M_{R}}.
\end{eqnarray}

For the signatures examined in this analysis,  the value of $M_{R}$ can have different interpretations. In the case of LQ$_3$ pair production, the LQ$_3$ corresponds to the particle $S$ from the above example, while $\chi$ is a neutrino. As a result, the characteristic scale $M_{\Delta}$ is an estimator of the LQ$_3$ mass. Similarly, for \sbottom pair production, $S$ refers to a \sbottom while $\chi$ is the LSP, generally a massive neutralino. In this case, $M_{\Delta}$ corresponds to the mass difference between the \sbottom and LSP.

As follows from the definitions above, $M_\mathrm{T}^{R}$ is expected to have a kinematic endpoint at the mass of the
new heavy particle, in a similar fashion to the transverse mass having
an edge at the particle mass (such as $M_\mathrm{T}$ in $\PW\to\ell\nu$ events).
Therefore, the $R$ variable is a measure of how well the missing transverse
momentum is aligned with respect to the visible momentum.  If the
missing momentum is completely back-to-back to the visible momentum, $R$
will be close to one.  On the other hand, if the momenta of the two
neutrinos or \neutralino largely cancel each other, $R$ will be
small. The distribution of $R$ for signal events will peak around
0.5, while for QCD multijet events it peaks at zero. These properties of
$R$ and $M_{R}$ motivate the kinematic requirements for the signal
selection and background reduction, which are discussed below.

Some differences between the kinematic distributions (such as the
transverse momenta of b jets) for LQ$_3$ production and \sbottom production may arise, if the mass of the \neutralino is substantial or even almost degenerate with the mass of the \sbottom. For a fixed
\sbottom mass the  $M_\Delta$ decreases as the \neutralino mass increases.
    In the case of an almost degenerate \neutralino and \sbottom, $\etmiss$ is relatively small and the jets are soft,
    resulting in an $M_{R}$ distribution shifted towards lower values,
    thus reducing the momentum of the \sbottom decays products and the sensitivity of the search.

\section{Data samples, triggers, and event selection}
\label{sec:sel}
 The analysis is designed using MC samples generated with
\PYTHIA (version 6.424)~\cite{Sjostrand:2006za} and \MADGRAPH~\cite{madgraph} (version 5.1.1.0),
and processed with a detailed simulation of the
CMS detector response based on \GEANTfour~\cite{G4}. Events with QCD
multijets, top quarks, and electroweak bosons are generated with \MADGRAPH interfaced with \PYTHIA tune Z2~\cite{Chatrchyan:2011id} for parton showering,
hadronization, and the underlying event description. Signal samples for
LQ$_3$ masses from 200 to 650\GeV, in steps of 50\GeV, are generated with
 \PYTHIA  tune D6T~\cite{Field:2008zz,Field:2010su}. The \sbottom pair production signal samples are generated with the  \PYTHIA
generator and processed with a detailed fast simulation of the CMS
detector response~\cite{CMS-DP-2010-039}. The scalar bottom quark signal samples are
generated with \sbottom masses from 100\GeV to 550\GeV in steps of 25\GeV, and
 \neutralino  masses from 50\GeV to 500\GeV
in steps of 25\GeV. The \sbottom samples are generated with the
assumption that the mass peak can be described by a Breit--Wigner
shape~\cite{Sjostrand:2006za}, but this assumption becomes imprecise
when the sparticles are close to degenerate. Samples where the
difference between the \sbottom mass and \neutralino mass is less than
50\GeV are therefore not generated. The simulated events are reweighted so that the distribution of number of  overlapping pp interactions per beam crossing (``pileup'') in the simulation matches that observed in data.

Events used in this search are collected by a set of online triggers. The first level (L1) of the CMS trigger system, composed of custom hardware processors, uses information from the calorimeters and muon detectors to select the most interesting events in a fixed time interval of less than 4\mus. The High Level Trigger (HLT) processor farm further decreases the event rate from around 100\unit{kHz} to around 300\unit{Hz}, before data storage. We employ three categories of triggers for this search: (i) hadronic razor triggers with
moderate/tight requirements on $R$ and $M_{R}$; (ii) muon razor triggers
with looser requirements on $R$ and $M_{R}$ and at least one muon in the
central part of the detector with $\pt>$ 10\GeV; and (iii)
electron razor triggers with the $R$ and $M_{R}$ requirements similar to those for muon razor triggers, and at
least one electron of $\pt>$ 10\GeV, satisfying loose isolation
criteria.  Events collected with the muon and electron razor triggers
are used to provide control regions for background studies, since the potential signal
contribution in these events is negligible. The search for the presence of a new physics signal is
performed in the events collected with the hadronic razor
triggers.

All events are required to have at least one good reconstructed
interaction vertex~\cite{TRK-10-005}.  Events containing calorimeter
noise, or  large $\etmiss$ due to instrumental effects (such as beam
halo or jets near non-functioning channels in the ECAL) are removed from the
analysis~\cite{METJINST}. The jets in the event, which are required to
have $|\eta| < 3.0$, are reconstructed from the calorimeter energy
deposits using the infrared-safe anti-\kt algorithm~\cite{antikt} with a distance parameter of 0.5, and are corrected for the
non-uniformity of the calorimeter response in energy and $\eta$ using
corrections derived from Monte Carlo and observed data~\cite{cms:jes}. The $\etmiss$
is reconstructed using the particle-flow algorithm, which identifies and reconstructs individually the particles produced in the collision, namely charged hadrons, photons, neutral hadrons, electrons, and muons~\cite{CMS-PAS-PFT-10-002}.

\subsection{Muon and electron identification and selection}
\label{sec:muons}

We select muon and electron candidates using a cut-based approach similar to the selection process used for the measurement of the inclusive $\PW$ and $\cPZ$ cross section~\cite{CMSWZxsections}.

We use the ``tight'' and ``loose'' muon identification criteria, and
all muons are required to have $\pt>20$\GeV. For loose muons, we
require that the muon candidate has at least 10 hits in the inner
tracker. For the tight muon we require in addition that the following
selections are met:

\begin{itemize}
\item at least one hit in the pixel detector;
\item impact parameter in the transverse plane $|d_{0}| < 0.2$\unit{cm};
\item $|\eta|<2.4$.
\end{itemize}

In addition, the tight muons satisfy a lepton isolation requirement $I_\text{comb}$
obtained by summing the $\pt$ of tracks and the energies of calorimetric
energy deposits in a cone of $\Delta R=\sqrt{(\Delta\eta)^2+(\Delta\phi)^2}<0.3$ around the lepton candidate,
excluding the candidate's $\pt$. We require the combined
isolation to be less than 15\% of the muon $\pt$.

 The selection requirements for prompt electrons are:

\begin{itemize}
\item $\pt>20$\GeV and $|\eta| < 2.5$;
\item combined isolation $I_\text{comb}<15$\% of electron $\pt$;
\item standard electron identification for barrel (endcap) electrons,
  defined as follows:
  \begin{itemize}
\item shape compatible with that of an electron, defined by a measure of the second moment of energy distribution among crystals $\sigma_{\eta\eta}< 0.012\,(0.031)$~\cite{CMSWZxsections};
  \item track-cluster matching in the $\phi$-direction, $\Delta\phi < 0.8\,(0.7)$;
  \item track-cluster matching in the $\eta$-direction, $\Delta\eta < 0.007\,(0.011)$.
  \end{itemize}
\end{itemize}

When the isolation requirements~\cite{CMSWZxsections} are applied to the
electron or tight muon candidates, the combined isolation $I_\text{comb}$ is
corrected for pileup dependence using the average energy density $\rho$ from other proton-proton collisions in the same beam crossing, calculated for each event~\cite{Cacciari:2007fd}.

\subsection{Identification of b jets}
\label{sec:btags}

Jets originating from a b quark are identified (``tagged'') by the
TCHE algorithm~\cite{CMS-PAS-BTV-11-004}. Selecting events with b-tagged
jets reduces the background from QCD multijet
events where mismeasured light-flavor jets cause large apparent
$\etmiss$. In the TCHE algorithm a jet is considered as b tagged if
there are at least two high-quality tracks within the jet, each with a
three-dimensional impact parameter (IP) significance IP/$\sigma_\mathrm{IP}$
larger than a given threshold (``operating point''). In this analysis we
use the ``medium'' operating point~\cite{CMS-PAS-BTV-11-004}. The
b-tagging efficiency ($\epsilon_\cPqb$) and mistag rate ($R_{\cPqb}$) have been measured up to $\pt$ = 670\GeV  and in the $\pt$ range 80--120\GeV are found to be $\epsilon_\cPqb=0.69\pm0.01$ and
$R_{\cPqb}=0.0286\pm0.0003$. In the following we refer to the sample
with two jets tagged by the medium TCHE tagger as the ``2b-tagged''
sample. A scale factor (per jet) of $0.95 \pm 0.02$ is applied to the to the MC simulation samples
to account for the observed differences in the b-tagging efficiency between the
simulation and data~\cite{CMS-PAS-BTV-11-004}.

\section{Search strategy}
\label{sec:ana}
Candidate signal events in this search contain a pair of b jets, large
$\etmiss$, and no isolated leptons. The main backgrounds that contribute to this final state originate from $\ttbar$+jets, HF
multijets, and $\PW/\cPZ$+HF jets events. Diboson production is included in the
total background estimation, but its contribution is small. Significant $\etmiss$ in multijet events derives from b quarks decaying
semileptonically or from jet energies being severely mismeasured.
Apart from the multijet background, the remaining backgrounds originate from processes with both genuine $\etmiss$ due to energetic
neutrinos and undetected charged leptons from vector boson decays.

Data sets collected with the razor
triggers are examined for the presence of a well-identified electron or
muon, as described in Section~\ref{sec:muons}. Based on the presence or
absence of such a lepton, the event is categorized into one of the three
disjoint event samples ({\it boxes}) referred to as the electron (ELE),
muon (MU), and hadronic (HAD) boxes.

These requirements define the inclusive {\it baseline} selection:

\begin{itemize}
\item MU box: events collected with muon razor triggers and containing one
  loose muon with $\pt > 20$\GeV, $M_{R}>400$\GeV and $R^2>0.14$.
\item ELE box: events collected with electron razor triggers and containing
  one loose electron with $\pt > 20$\GeV, $M_{R}>400$\GeV and $R^2>0.14$.
\item HAD box: events collected with hadronic razor triggers and not
  satisfying any other box requirements, and with $M_{R}>400$\GeV and $R^2>0.2$.
\end{itemize}

We also require that there are at least two jets above 60\GeV in each
event, to ensure that the trigger is fully efficient for our selected events. In order to study and estimate the
background contributions in the HAD box, we treat muons and electrons in
the MU and ELE boxes as neutrinos, i.e. the lepton 4-vector is used to
recalculate the $\etmiss$ vector and the $R$ variable is recomputed.
This procedure generates the kinematic properties of the background
events in the HAD box, using events from the MU and ELE boxes that, because of the presence of the leptons, are free of the signals
relevant to this analysis.

The distributions of the discriminating variables $R$ and $M_{R}$ for the main
backgrounds (heavy-flavor multijets and $\ttbar$) are estimated from observed
data. Events in the MU box are used to extract the probability
density functions (PDFs) describing the behavior of the $R$ and $M_{R}$
shapes for each process of interest. For the $\PW/\cPZ+$HF-jets and diboson
backgrounds we use heavy-flavor-enriched \MADGRAPH simulation
samples to get the shape prediction. The procedure to extract the
background shapes is described in detail in Section~\ref{sec:bkg}, and
the samples used are summarized in Table~\ref{tab:boxes}.

To predict the SM background normalizations in the signal region we
adopt the following strategy. The events in the ELE and HAD boxes are
split into two exclusive categories:

\begin{itemize}
\item sideband: events with $400<M_{R}<600$\GeV and $0.2<R^2<0.25$;
\item high $R^2$: events with $M_{R}>400$\GeV and $R^2>0.25$.
\end{itemize}

The 2b-tagged high-$R^2$ events in the HAD box define the signal
search region.  The normalizations of the SM backgrounds in the signal
region are obtained through a two-step procedure:

\begin{itemize}
\item the SM processes are normalized according to their theoretical
  cross sections, except for $\ttbar$ where the measured CMS cross
  section~\cite{Chatrchyan:2011nb} is used;
\item the total background prediction in the high-$R^2$ region is
  multiplied by a scale factor ($f_{R^2}$) to correct for imperfect
  knowledge of the multijet production cross section.
\end{itemize}

The scale factor is derived from events in the sideband, and is defined as $f_{R^2}=N_\text{exp}/N_\text{obs}$, where
$N_\text{exp}$ is obtained using the background PDF normalized to their
individual cross sections; and $N_\text{obs}$ is the number of observed events.

In order to avoid potential bias in the search, before analyzing the
events in the HAD box signal region, we test our understanding of the SM
background estimation procedure in control regions, using the MU and ELE
boxes. This is done by comparing the background shapes derived from the
MU box to the observed data in the ELE box (removing the leptons from the
reconstruction to emulate $\etmiss$ in each case). To ensure that both the
shapes and normalizations of the background components describe the
observed events, the procedure to be used in the HAD box (see Table~\ref{tab:boxes}
below) is first employed and tested in the ELE box (Sec.~\ref{sec:elebox}). Events in the ELE sideband are used to obtain the scale factor $f_{R^2,\,\mathrm{ELE}}$ which is used to test the background prediction in high $R^2$ ELE box. Once the procedure is validated in the ELE box, the $f_{R^2,\,\mathrm{HAD}}$ is derived from  events in the sideband of the HAD box, and is used to predict the normalization of the backgrounds in the signal region.

\begin{table}[htdp]
  \topcaption{Summary of   samples used in the search, with a short description of their specific purpose. Events in all samples are required to have $M_{R}>400$\GeV and to include two b-tagged jets. The selections on $R^2$ listed in the table are applied after recalculating $\etmiss$ and $R$ for events in which charged leptons are treated as neutrinos.  The definitions of muons ($\mu$) and electrons ($\Pe$) are discussed in Section~\ref{sec:muons}.}
  \begin{center}
    \begin{tabular}{cccc}
      \hline
      Sample &   $R^2$ cut & Leptons & Comment \\ \hline\hline
      $\PW/\cPZ$ MC  & $R^2>0.07$ & tight $\mu$ & shape of $\PW/\cPZ$+HF jets \\
      MU   & $R^2>0.14$ & tight $\mu$    & shape of $\ttbar$+jets \\
      MU  & $R^2>0.14$ & loose $\mu$   & shape of HF multijets \\
      ELE  & $0.2<R^2<0.25$ & tight $\Pe$  &  $M_{R}<600$, sideband to extract $f_{R^2,\,\mathrm{ELE}}$\\
      ELE  & $R^2>0.25$ & tight $\Pe$     & ELE ``signal-like'' control region \\
      HAD& $0.2<R^2<0.25$ & veto leptons     & $M_{R}<600$, sideband to extract $f_{R^2,\,\mathrm{HAD}}$ \\
      HAD& $R^2>0.25$ & veto leptons  & signal box, search for signal\\

      \hline
    \end{tabular}
  \end{center}
  \label{tab:boxes}
\end{table}

 \section{Background estimation}
 \label{sec:bkg}
 In both simulation and observed data, the distributions of SM background events
have been shown to have a simple exponential dependence on the razor
variables $R$ and $M_{R}$ over a large fraction of the $R^2$-$M_{R}$
plane~\cite{rogan,SUS-10-009}. The shape of the $M_{R}$ tail is well-described by two exponentials
with slope parameters $S_i$ ($i=1,2$), where
each $S_i$ depends linearly on the $R^2$ selection threshold ($R_\text{min}^2$):
$S_i=A_i+B_i\times R_\text{min}^2$.

We construct a simultaneous fit across different $R$ bins, where the
$M_{R}$ distribution is fitted for each value of the $R^2$ threshold to extract
the $A_i$ and $B_i$ parameters. The simultaneous fit allows one to fully
exploit the correlations between the fit parameters and therefore (\textit{i}) to get a better estimate on the uncertainty of the $A_i$ and $B_i$
parameters, and (\textit{ii}) to ensure that the PDF obtained from the fit
can be used in regions with various $R^2$ thresholds. The functional form used in the fit for a fixed value of the $R$ threshold is:

\begin{equation}
  \label{eq:fit}
  F(M_{R})=\re^{-(A_1+B_1\times R_\text{min}^2)M_{R}}+f\times \re^{-(A_2+B_2\times R_\text{min}^2)M_{R}},
\end{equation}

where $f$, the relative amplitude of the second exponent, is extracted
from the fit. The values of the shape parameters that maximize the likelihood in the fits,
along with the corresponding covariance matrix, are used to define the
background model and the uncertainty associated with it. Therefore, if a
pure sample of a given process is selected, the PDF describing the
behavior of the $R$ and $M_{R}$ shapes of a given process can be
extracted.

The fits are performed using the \textsc{RooFit}
toolkit~\cite{verkerke-2003}. The background PDFs are then used to
generate pseudoexperiments, to evaluate the effects of
systematic uncertainties on the event yields, as described below in
Section~\ref{sec:systematics}.

\subsection{The \texorpdfstring{$\PW/\cPZ$+jets}{W/Z+jets} background \label{bg-WZ}}
\label{sec:WZ}
Owing to the lack of a high-purity data sample enriched in events with
$\PW/\cPZ$+two heavy-flavor jets, we estimate the shape of the $\PW/\cPZ$+jets
background using MC simulated events. A selection of events in the observed data
whose jets fail to be b-tagged could provide a sample
enriched in $\PW$+light flavor jets. However, because of the
b-tagging efficiency on the jet $\pt$ ~\cite{CMS-PAS-BTV-11-004}, the PDF extracted from these events does not provide
a sufficiently accurate model for $\PW/\cPZ$+b jets events. Therefore, we estimate
the shape of the $\PW/\cPZ$+jets background using simulated events generated
with the \MADGRAPH event generator interfaced with \PYTHIA, which
were found to give an adequate description of CMS
observed data~\cite{CMS-PAS-EWK-11-017, Chatrchyan:2012vr}. Residual deficiencies of this MC simulation-based background
modeling are accounted for in the extraction of the $\ttbar$
background estimate from observed data, as described in the
Section~\ref{sec:top}. The overall normalization of this background is
determined using the observed events in the sideband region of the HAD box.

\begin{figure}[ht!]
  \begin{center}
    \includegraphics[width=0.495\textwidth]{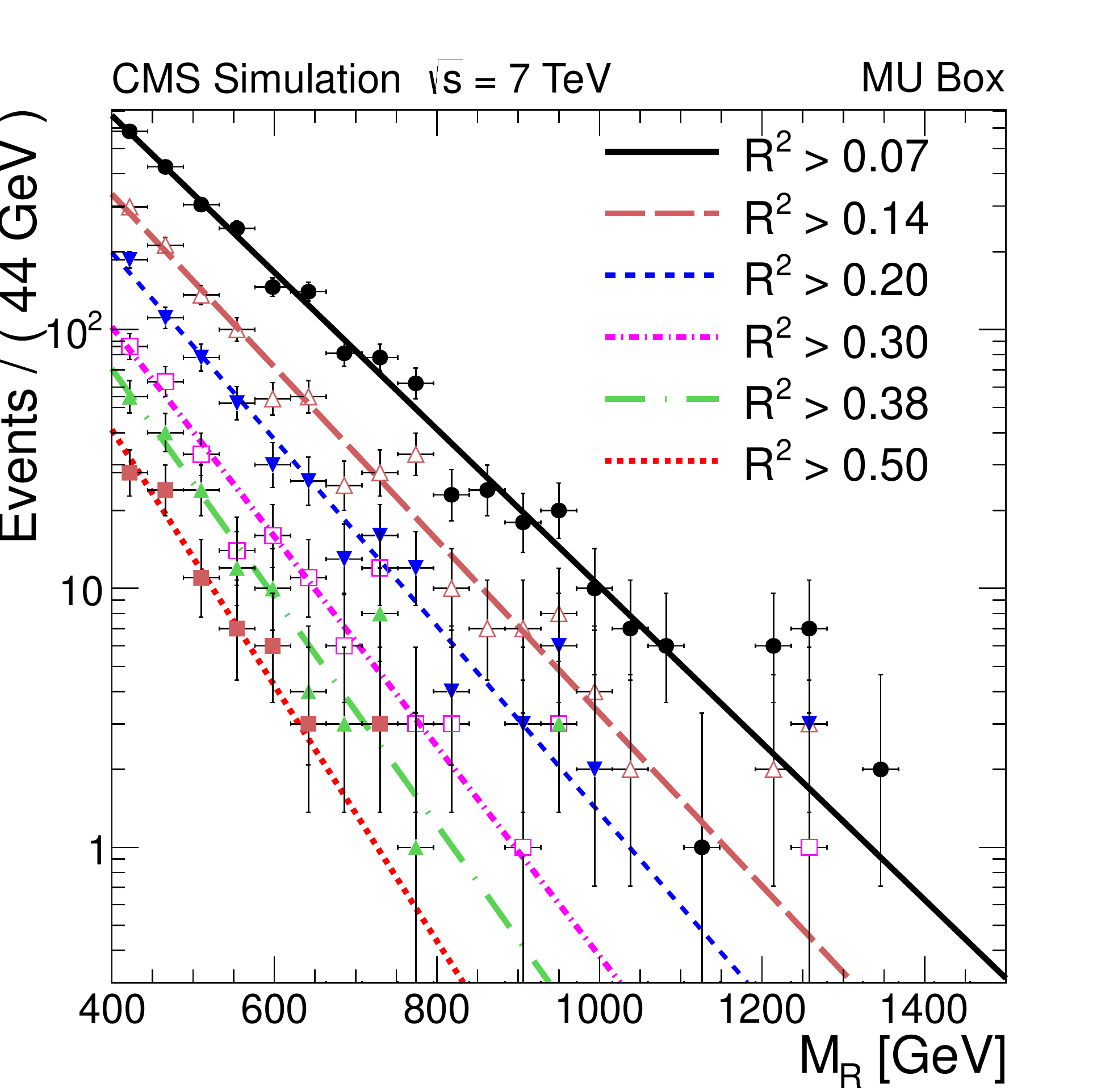}
    \caption{$M_{R}$ distributions for different values of the
      $R^2$ threshold for events passing the MU box selections in the
      $\PW/\cPZ$+jets MC simulation. The results of the fits (lines)
      are overlaid with the $M_{R}$ distributions from the MC simulation (markers).}
    \label{fig:WZjetsFits}
  \end{center}
\end{figure}

We perform an unbinned fit of the $\PW/\cPZ$+jets $M_{R}$ distribution in simulated
 events passing the MU box selections with 2b-tagged events,
using the sum of two separate exponential terms, as shown in Eq.~(\ref{eq:fit}). The fit
allows us to obtain a parametric description of the background that is
later used in the derivation of the remaining backgrounds, and it also
permits the extrapolation of the prediction into the region of higher
$R$ and $M_{R}$ values. The fit is performed in the region $M_{R}>400$\GeV and is
binned in values of $R^2$ as shown in Fig.~\ref{fig:WZjetsFits}.
The fit to the simulated data, which provides a good description of the $M_{R}$ distribution, is used as the PDF  to estimate the $\PW/\cPZ$+b jets background in the signal box.

\subsection{\texorpdfstring{$\ttbar$+jets}{t t-bar + jets} background estimation\label{sec:top}}
\label{sec:top}
\begin{figure}[ht!]
  \begin{center}
    \includegraphics[width=0.495\textwidth]{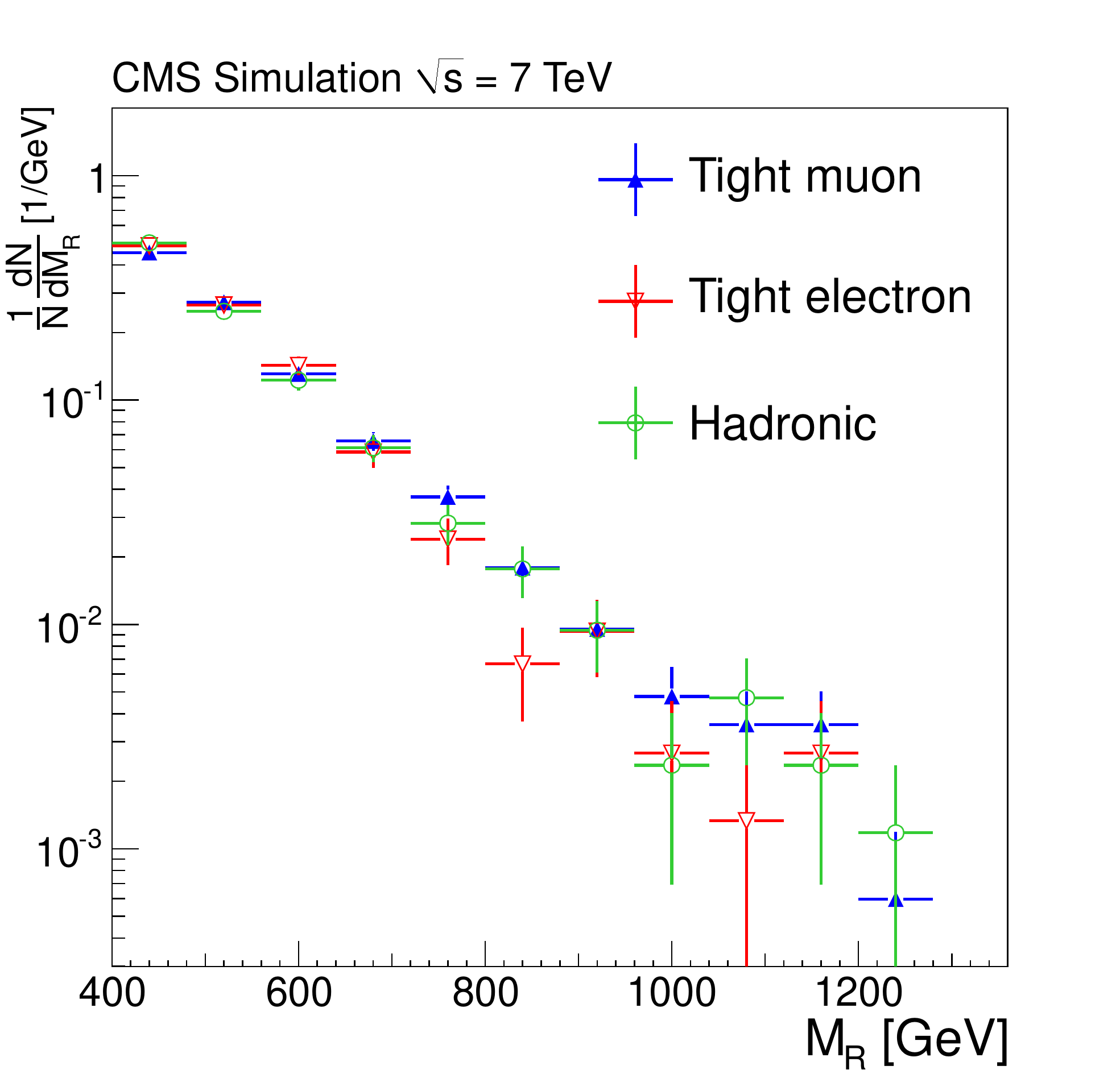}
    \includegraphics[width=0.495\textwidth]{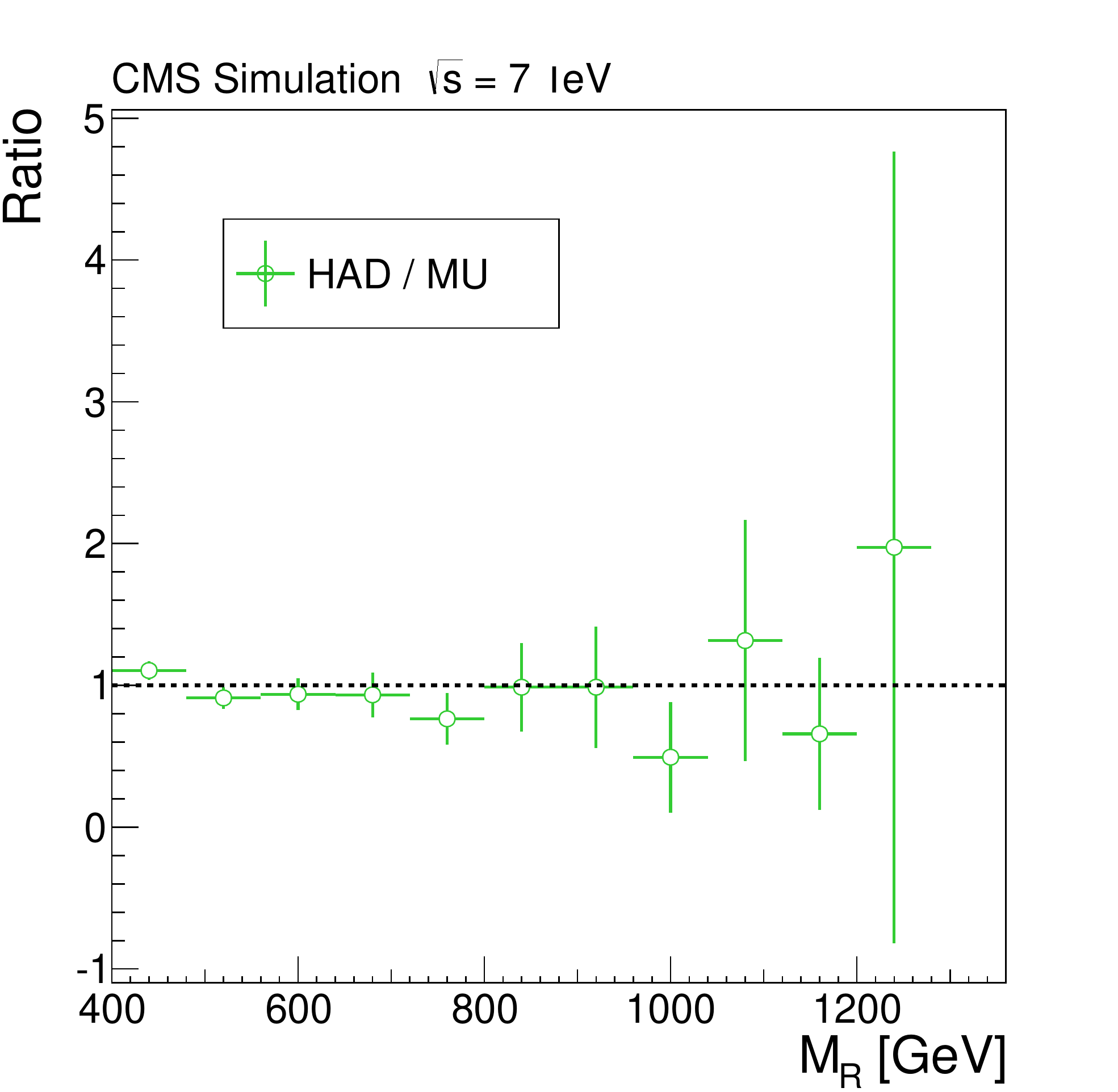}
    \caption{The $M_{R}$ distributions (left) in $\ttbar$ MC simulated events selected with
      either tight MU, tight ELE and HAD requirements, and (right) the ratio of the number of events selected with the HAD or tight MU selections, as
      a function of $M_{R}$. }
    \label{fig:ttbarInMC}
  \end{center}
\end{figure}

\begin{figure}[ht!]
  \begin{center}
    \includegraphics[width=0.495\textwidth]{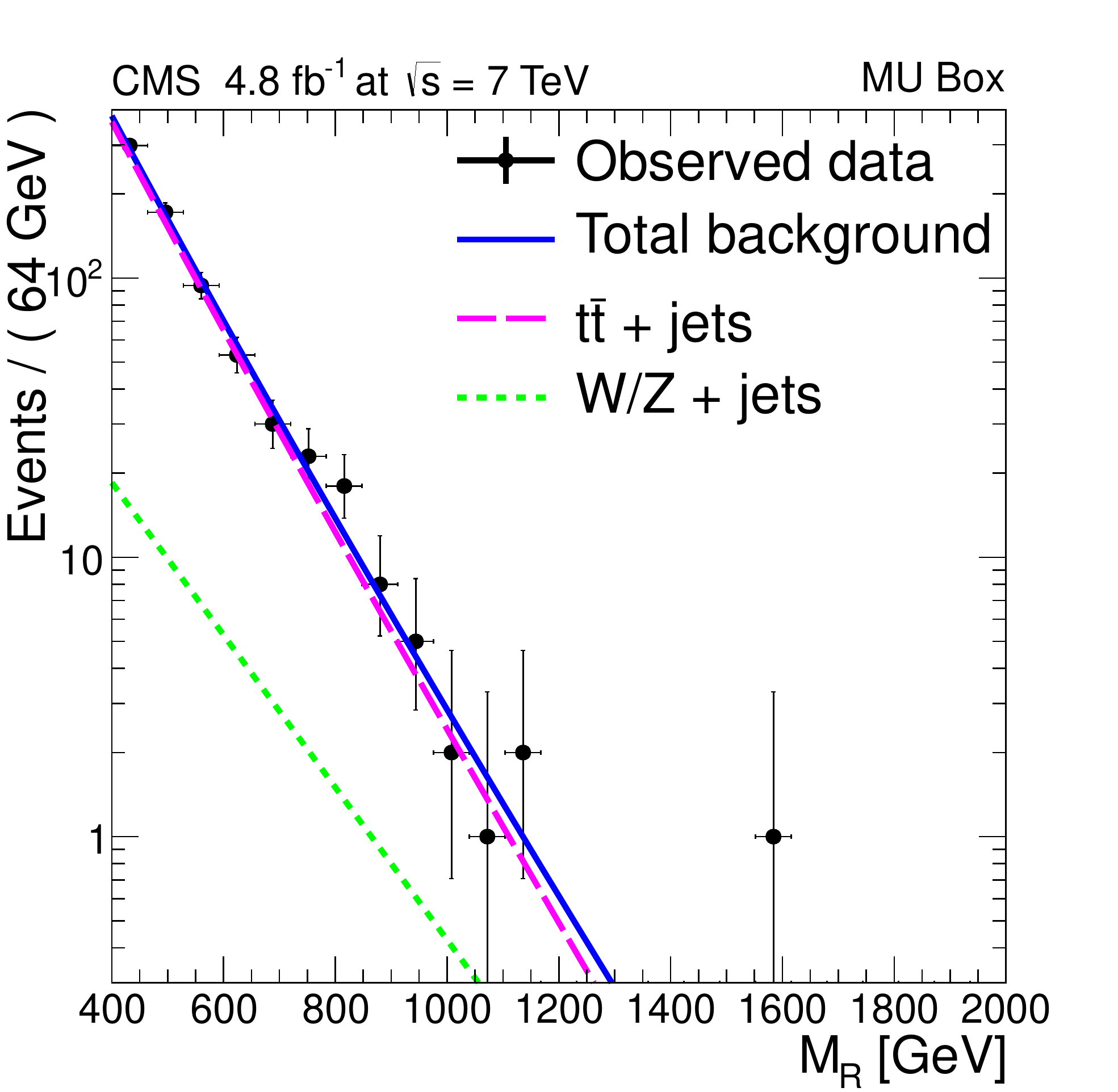}
    \includegraphics[width=0.495\textwidth]{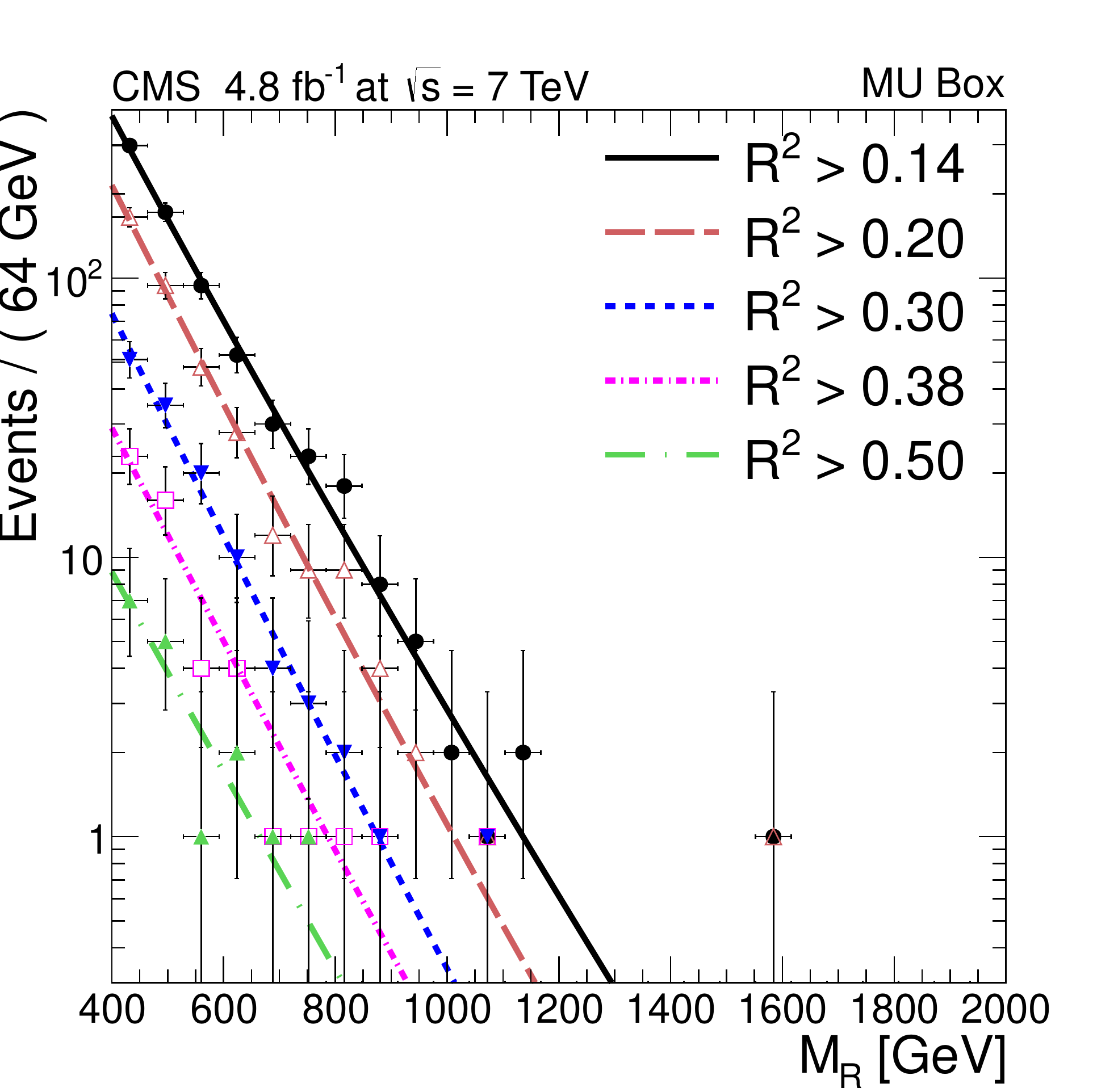}
    \caption{The result of the fit of the $M_{R}$ distributions (lines) compared to MU box observed events with $R^2>0.14$ (left); individual
      background contributions are not stacked. On the right are shown the $M_{R}$ distributions for different values of the
      $R^2$ threshold (right) in 2b-tagged events of the MU box with a tight muon; the results of the fits (lines) are overlaid with the
      observed distributions (markers).}
    \label{fig:tbarInMUStarL}
  \end{center}
\end{figure}

We estimate the $\ttbar$  background from the MU box, using 2b-tagged
 events in collision data (Section~\ref{sec:btags}) and requiring the presence of a
muon passing the tight identification requirements
(Section~\ref{sec:muons}). Based on comparisons with the MC simulation,
approximately 90\% of the events in this sample are $\ttbar$. We
find empirically from MC simulation studies that the shape of the $M_{R}$
distribution in both the tightly selected MU box and in the HAD
box is very similar, as can be seen in Fig.~\ref{fig:ttbarInMC}. We therefore use
the shape derived from the 2b-tagged sample to predict the $\ttbar$
background in the signal region. Additionally, because of a non-negligible contribution of $\PW/\cPZ$+HF events in this sample, the imperfections in the $\PW/\cPZ+\text{jets}$ background modeling in the simulation are absorbed into the $\ttbar$ background prediction. In order to derive the $\ttbar$
shape, we constrain the $\PW/\cPZ+\text{jets}$ shape to that obtained from the MC
simulation~(Section~\ref{bg-WZ}). We find that a two-exponential
function provides a good fit to the observed data in the MU box, as shown in
Fig.~\ref{fig:tbarInMUStarL}.

\subsection{Multijet background}
\label{sec:multijet}

The remaining backgrounds that contribute significantly to the interesting region of high $R^2$ originate from heavy-flavor enriched multijet
production. We use events with a loose muon in the MU box to derive the
multijets background PDF. According to the MC simulation, this sample is
composed 45\% of top events, 5\% of $\PW/\cPZ$+b jet events, and 50\% of
multijet events.

\begin{figure}[ht!]
  \begin{center}
    \includegraphics[width=0.495\textwidth]{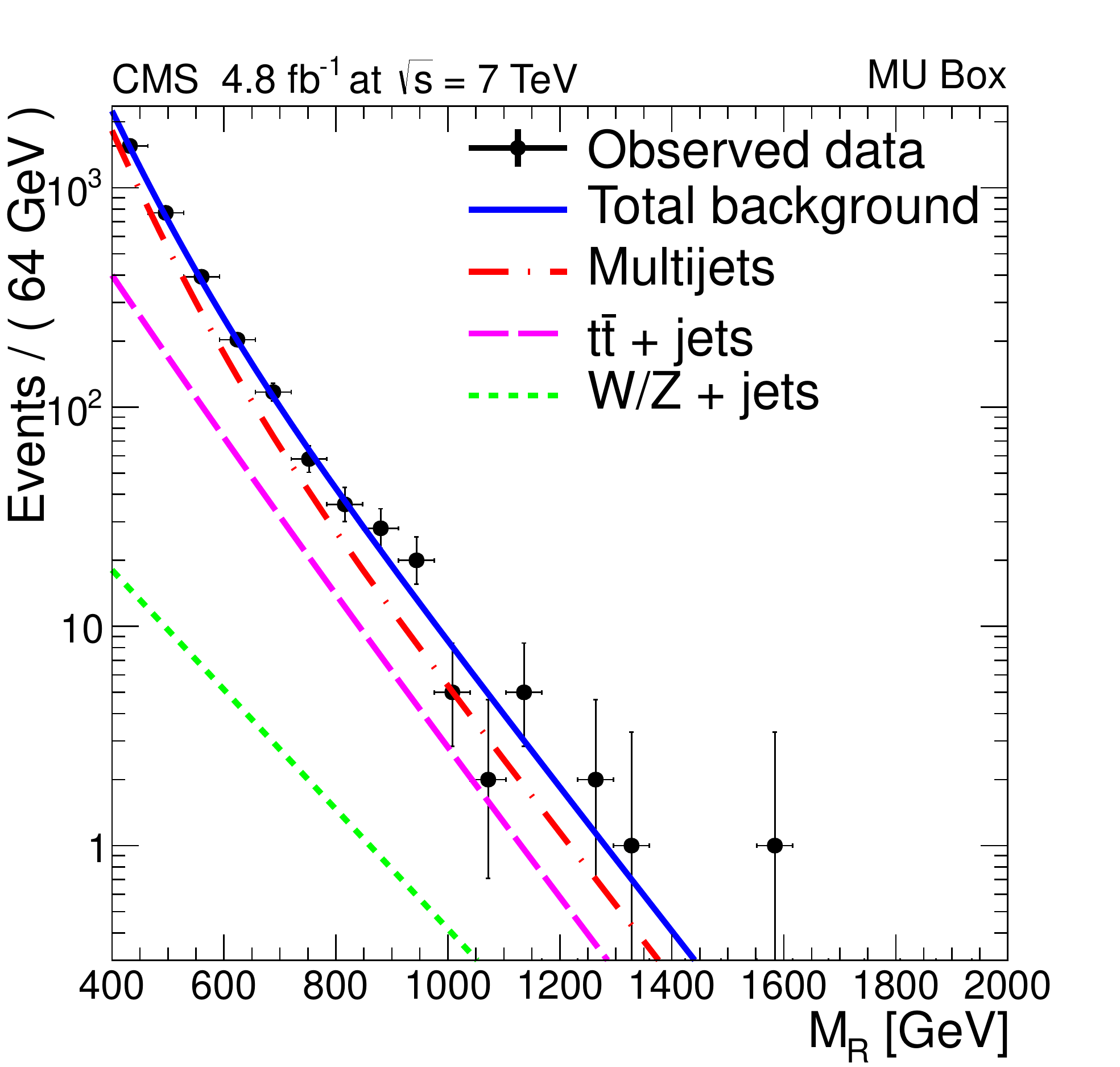}
    \includegraphics[width=0.495\textwidth]{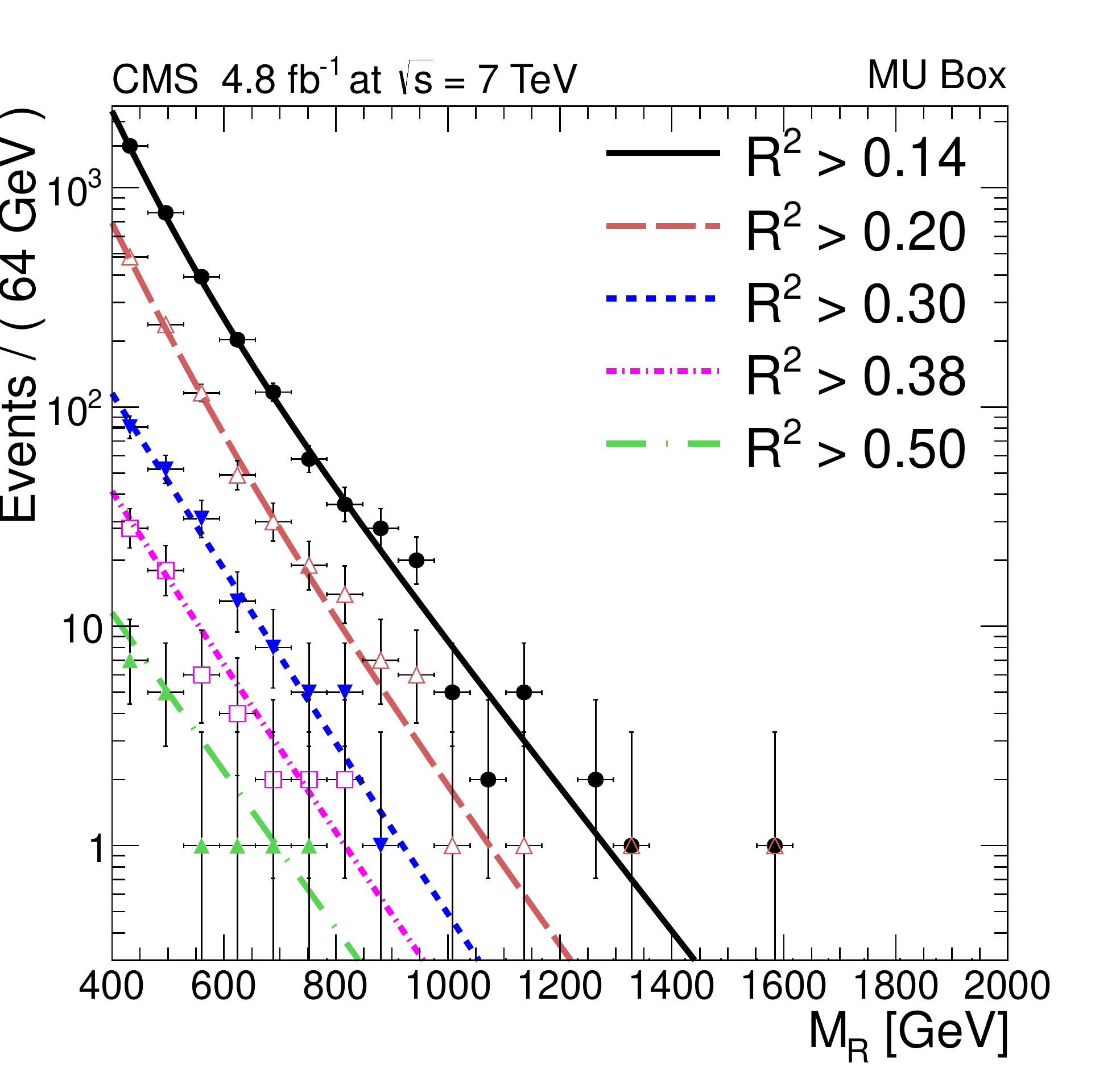}
    \caption{The result of the fit of the $M_{R}$ distributions (lines) compared to the MU box observed data for events with $R^2>0.14$ (left); individual contributions of backgrounds are not stacked. On the right are shown the $M_{R}$ distributions for different values of the $R^2$ threshold (right) in 2b-tagged events of the MU box with a loose
      muon; the results of the fits (lines) are overlaid with the observed distributions (markers).}
    \label{fig:multijets}
  \end{center}
\end{figure}

We proceed to perform the fits, for which the contributions from $\PW/\cPZ$+b
jets and $\ttbar$ backgrounds are fixed to the PDFs described in
Sections~\ref{sec:WZ} and \ref{sec:top}. Based on simulation studies it is found that the parameters of the second component in the fit
function ($A_2$ and $B_2$ in Eq.~(\ref{eq:fit})) are nearly idenical for the multijet and the $\ttbar$+jets background
processes. In order to better constrain the multijet fit, the parameters
of the second component are set equal to those from the
observed events for $\ttbar$+jets while the parameters of the first
component of the multijet PDF are left free. The results of the fit in
the 2b-tagged MU box are displayed in Fig.~\ref{fig:multijets}, where
we find good agreement between the fit results and observed data.

\subsection{Systematic uncertainties}
\label{sec:systematics}

For the backgrounds estimated from observed events, the uncertainty in the
total yield arises from the uncertainties (statistical and systematic)
in the fit parameters in Eq.~(\ref{eq:fit}).  We estimate these
uncertainties by varying the $R^2$ threshold values (by ${\pm}5$\%),
thus arriving at a new set of $A_i$ and $B_i$ parameters describing the
background PDF. The maximum difference observed between the experimental data and the simulated data in the MU box with tight and loose muon selections is then used as the
uncertainty on the shape parameters. This procedure results in a 10\%
uncertainty in the $A_i$ values, and 40\% in the $B_i$ values. We also
tested the stability of the fits by varying the initial parameters used
to start the fit by ${\pm}50$\%, and found that this variation results in stable
solutions, returning the same central value for the $A_i$ and $B_i$
parameters.

We generate an ensemble of pseudoexperiments, based on the fit
results in the MU box. From each pseudoexperiment a new set of values for the
parameters is then obtained, with the corresponding uncertainties,
and we use the associated PDF results to predict
the background yield. The ensemble of pseudoexperiments thus provides a distribution
of the expected background yield in the signal regions, with its
corresponding uncertainty. This procedure allows us to correctly
propagate the systematic uncertainty in the background shape into the
prediction of the background. To account for the normalization
uncertainty we propagate the uncertainty in the $f_{R^2}$ introduced in
Section~\ref{sec:ana} to the prediction of background yields in the signal region from control samples in observed events.

The effect of the jet energy scale (JES) and jet energy resolution (JER)
 uncertainties on the $\PW/\cPZ$+jets background estimate and the signal model yields from simulation are taken into account. These effects are evaluated by repeating
the extraction of all background PDFs by first varying the JES/JER by
plus or minus one standard deviation in the $\PW/\cPZ$+jet background model, and
recalculating the $\etmiss$ and $R$. These variations correspond to uncertainties as large as 3\% in the
selection efficiency. We then re-derive the background model PDFs from observed data in the MU box, using the newly obtained $\PW/\cPZ$+HF jets model. The new set of PDFs with their corresponding covariance
matrices then serve as an alternative background model.

We apply a scale factor of about 0.95, that is weakly dependent on jet $\pt$, to account for an observed difference in tagging efficiency between data and simulation. The uncertainty in the scale factor varies from 0.03 to 0.05
for jets with $\pt$ from 30 to 670\GeV, and is 0.10 for b jets with
$\pt> 670\GeV$. These uncertainties are measured using a dijet sample
with high b-jet purity, as detailed in Ref.~\cite{CMS-PAS-BTV-11-004}.

The uncertainty in the \sbottom acceptance due to uncertainties in the parton distribution functions is calculated using the recommendation from
the PDF4LHC group~\cite{Botje:2011sn}. The parton distribution function and the $\alpha_s$ variations of next-to-leading (NLO) order in the MSTW2008~\cite{Martin:2009iq}, CTEQ6.6~\cite{Nadolsky:2008zw}, and NNPDF2.0~\cite{Ball:2010de} sets were taken into account and their impact on the signal cross sections was compared with the calculation with CTEQ6L1~\cite{PDFLQ3} that was used in the simulation of the signal samples. From these three
 sets we evaluate an upper and lower bound on the signal
efficiency for each pair of assumed \sbottom and \neutralino masses, and half of the difference between the two bounds is used as an estimate of the uncertainty. The theoretical cross section of LQ$_3$ production has been calculated
using CTEQ6L1 and CTEQ6M~\cite{PDFLQ3}  at NLO, and the uncertainty in the prediction of the cross section was estimated by repeating the calculation using the NLO MRST2002 parametrization~\cite{Martin:2002aw}. This uncertainty was found to vary from 3.5 to 25\% for leptoquarks in the mass range considered in this analysis ~\cite{Kramer:2004df}.

The systematic uncertainty to the luminosity measurement is taken to be 2.2\%~\cite{CMS-PAS-SMP-12-008}, which is correlated among all signal channels and the background estimates that are derived from simulations. The uncertainty in trigger efficiency is estimated using a set of prescaled razor triggers with low thresholds, and is found to be 2\% for events in the HAD box, and 3\% for events in the MU and ELE boxes.

\subsection{ELE control region}
\label{sec:elebox}

In order to check that our background shape modeling indeed predicts the
observed data adequately, we use the PDFs obtained in the steps
described above (Sections~\ref{sec:WZ}-\ref{sec:multijet}) in an
orthogonal sample in the 2b-tagged ELE box with a tight electron
selection, i.e. the sample with a well-identified electron, which is
then treated as a neutrino.  This signal-depleted sample provides an
independent cross-check of our background modeling, and covers the same
region in $R$ and $M_{R}$ as the HAD box. Additionally, based on MC simulation
studies, the composition of the tight ELE sample in observed events is similar to that
of the HAD sample, consisting of approximately 85\% $\ttbar$, 5\%
 $\PW/\cPZ$+HF jets, and 10\%  multijet events. For comparison, the HAD
sample is expected to contain approximately 70\%, 5\%, and 25\% of the
respective backgrounds.

Using the background model PDFs obtained from the fits, we derive the
distribution of the expected shapes in the ELE box using
pseudoexperiments. In order to correctly account for correlations and
uncertainties in the parameters describing the background model, the
shape parameters used to generate each pseudoexperiment data set are
sampled from the covariance matrix returned by the fit. The actual
number of events in each dataset is then drawn from a Poisson
distribution centered on the yield returned by the covariance-matrix
sampling. For each pseudoexperiment dataset, the number of events in
the sideband and in the high-$R^2$ region is found. We then obtain the
scale factor $f_{R^2,\,\mathrm{ELE}}=0.87\pm0.14$ from the sideband region, which is
used to predict the overall yield of background events in the high $R^2$
region of the ELE box.

\begin{figure}[ht!]
  \begin{center}
    \includegraphics[width=0.495\textwidth]{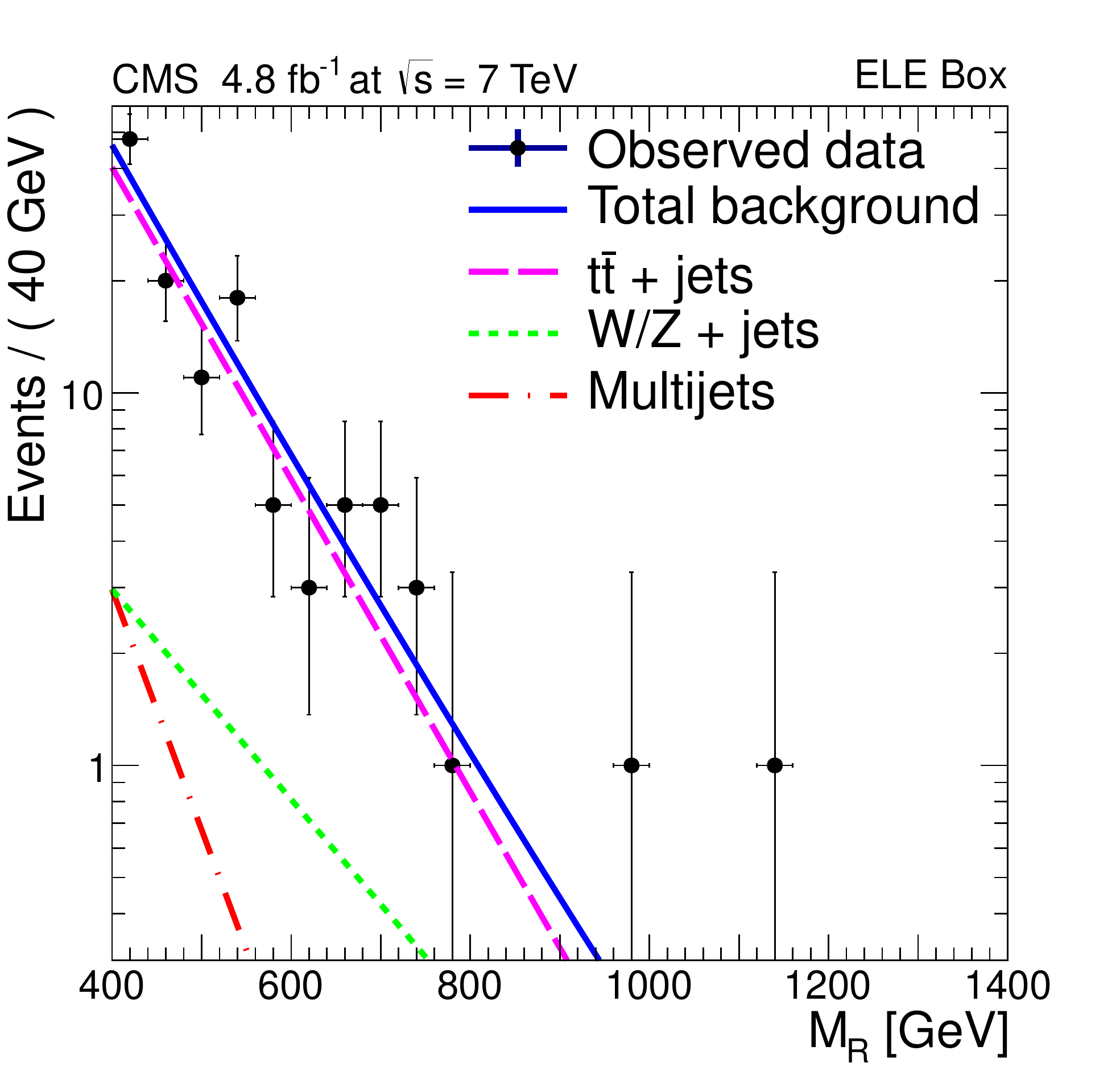}
    \caption{The $M_{R}$ distribution for observed data in the 2b-tagged ELE
      box  for events with $R^2>0.25$ compared to the prediction. The background model
      derived from the MU box is used to predict the $M_{R}$ shapes of the
      background processes.  The individual contributions are not
      stacked.}
    \label{fig:ELEStartCR}
  \end{center}
\end{figure}

The comparison of the predicted $M_{R}$ distribution with the observed events in the
ELE box is shown in Figure~\ref{fig:ELEStartCR}, and the background model is found to predict the
observed data adequately. We also test our
ability to correctly predict the yields of SM backgrounds using the
scale factor mentioned above. The results are summarized in
Table~\ref{tab:ELEstarYields}. Total background yield in the sideband is normalized to the number of observed data events in the sideband, in order to derive the scale factor  $f_{R^2,\,\mathrm{ELE}}$,  as described in Section~\ref{sec:ana}. The uncertainties in the background yields shown here represent systematic uncertainties that are estimated by varying the parameters $A_i$ and $B_i$, as described in Section~\ref{sec:systematics}.
As can be seen in this comparison, the $f_{R^2,\,\mathrm{ELE}}$
obtained from the sideband allows one to predict the overall
normalization of the 2b-tagged sample.

\begin{table}[!ht]
\begin{center}
\topcaption{Comparison of the yields in the ELE box. The sideband here refers to 2b-tagged events in the ELE box with $400<M_{R}<600$\GeV and $0.2<R^2<0.25$, while ``signal-like'' refers 2b-tagged
events with $M_{R}>400$\GeV and $R>0.25$. The scale factor derived in the sideband ($f_{R^2,\,\mathrm{ELE}}=0.87\pm0.14$) is used to normalize
the background yield in the signal-like region (third column), and the  uncertainty on the $f_{R^2,\,\mathrm{ELE}}$ is propagated into the total background yield.
\label{tab:ELEstarYields} }

\begin{tabular}{ccc}\hline
                                                                                   & Sideband & Signal-like\\
                                                                                   \hline
Multijets                                                                    & $12.5	\pm	1.9$ & $10	\pm	11$ \\
W/Z+jets                                                                   & $3.6	\pm	1.9$ & $8.8	\pm	2.8$ \\
$\ttbar$+jets                                        & $58.8	\pm	7.7$ & $118.4	\pm	9.8$ \\
Other backgrounds                                               & $0	\pm	0$  & $0.6	\pm	1.0$ \\
\hline
$f_{R^2,\,\mathrm{ELE}}$                                    &  \multicolumn{2}{c}{$0.87\pm0.14$} \\
\hline
Total background                                                   & $65	\pm	13$  & $119	\pm	23$ \\
\hline\hline
Observed data                                                        &  65 & 121\\
\hline

\end{tabular}
\end{center}
\end{table}

We perform another check to test whether the $R^2$-dependence is
well-described by our background model. This check is needed since in the
final signal region we have several signal boxes, each optimized for
different signal mass hypotheses. In order to increase the sensitivity
for higher masses, a tighter selection on $R^2$ is imposed to reduce the
backgrounds further, while keeping the signal efficiency high. In
order to ensure that our background model adequately describes
observed data with higher $R^2$ thresholds, we perform the same procedure in the ELE
box. The results are summarized in Table~\ref{tab:ELEboxcounts}. Here, we use
the same $f_{R^2,\,\mathrm{ELE}}$  derived from the sideband. As can be seen from these
results, this model correctly predicts the total yields for higher $R^2$ boxes.

\begin{table}[ht]
\begin{center}
\topcaption{Expected and observed yields in the 2b-tagged ELE box for $R^2$ selections and a fixed requirement $M_{R}>400$\GeV. The quoted
  uncertainties on the expected number of events include statistical and
  systematic uncertainties, and the uncertainty on the
  scale factor $f_{R^2,\,\mathrm{ELE}}$.}
{
\begin{tabular}{ccc}\hline
$R^2$  Cut    & Expected yields  & Observed yields \\
\hline
${>}0.25$      & 	$119	\pm	23$	&	121	\\
\hline
0.25--0.30	&	$51	\pm	17$	&	48	\\
0.30--0.35	&	$30	\pm   10$	&	26	\\
0.35--0.38	&	$9.9   	\pm	5.2$	&	11	\\
0.38--0.42	&	$11.5      	\pm	5.0$	&	11	\\
${>}0.42$	&	$16.8	\pm	4.8$	&	25	\\
\hline
\end{tabular}
}

\label{tab:ELEboxcounts}
\end{center}
\end{table}

 \section{Results}
 \label{sec:results}
\begin{figure}[ht!]
  \begin{center}
    \includegraphics[width=0.495\textwidth]{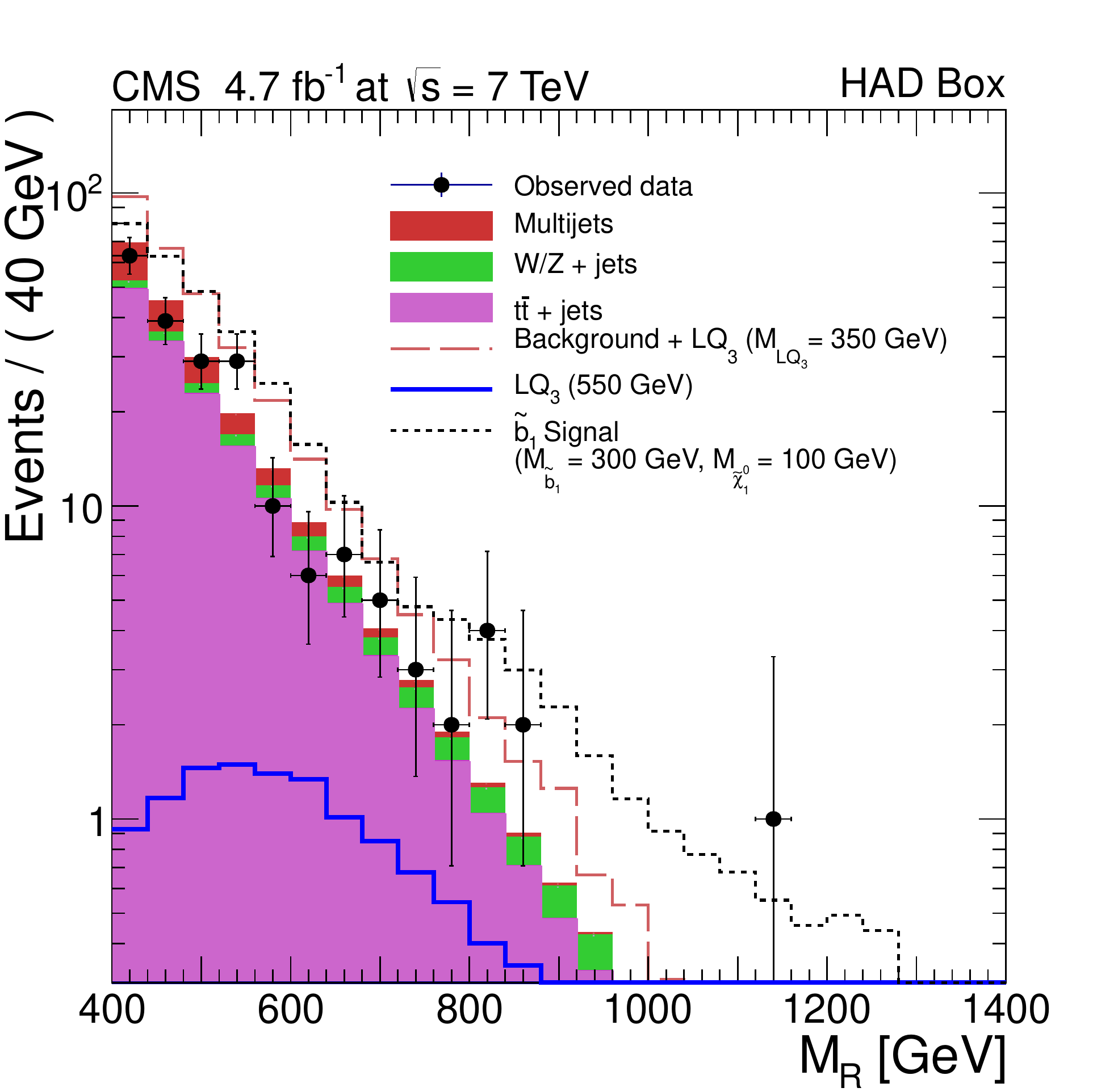}
    \includegraphics[width=0.495\textwidth]{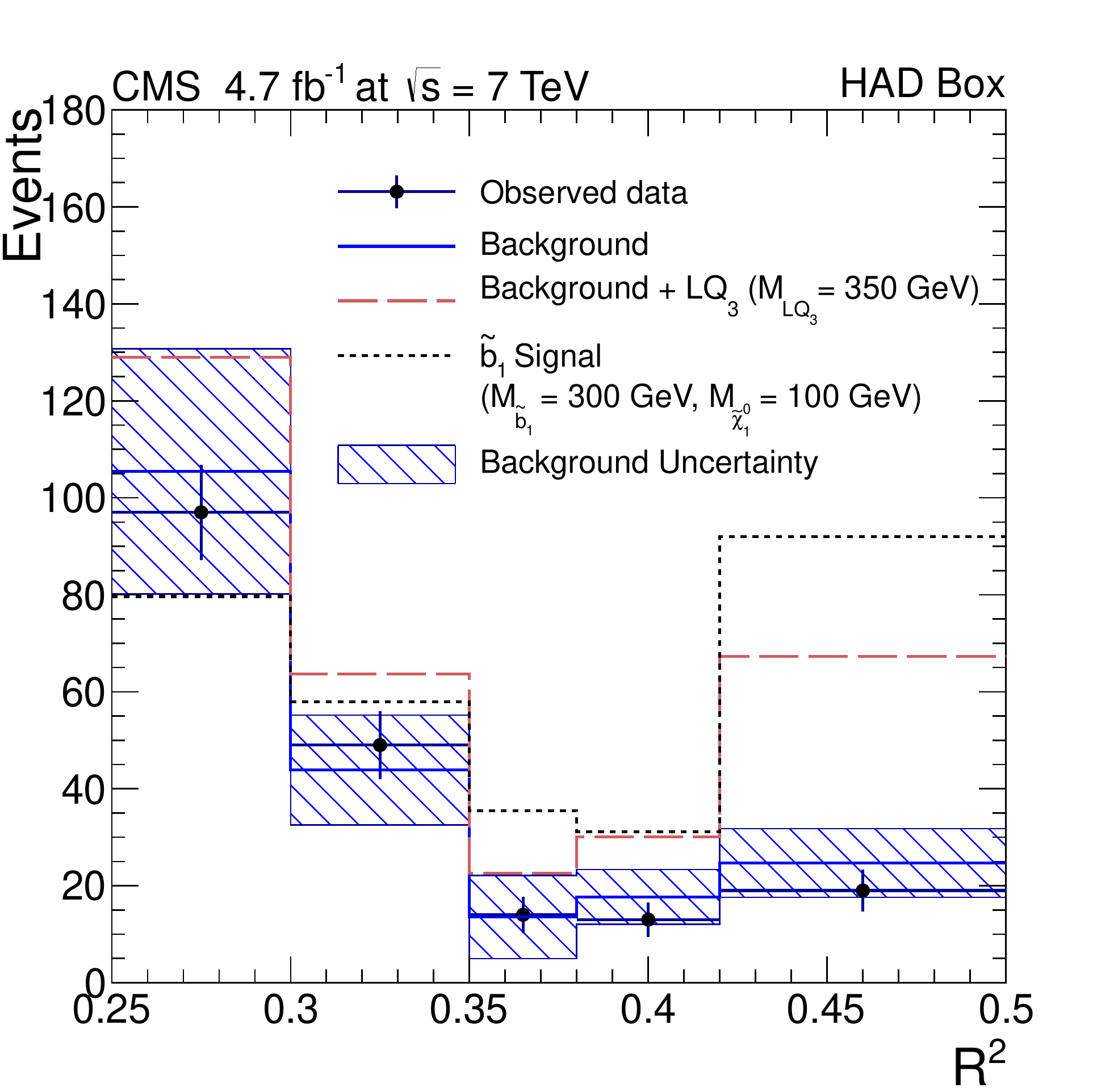}
    \caption{Comparison of the background prediction with the data
      observed in the 2b-tagged sample in the HAD signal box for the
      $M_{R}$ (left) and $R$ (right) distributions. The expected
      contributions from LQ$_3$ and \sbottom signal events with various mass
      hypotheses are also shown.}
    \label{fig:signalbox}
  \end{center}
\end{figure}

We search for LQ$_3$ and \sbottom signals in the HAD box data sample using the background PDFs obtained
from the MU box (Sections~\ref{sec:WZ}-\ref{sec:multijet}). The predicted background yields and their uncertainties  are
summarized in Table~\ref{tab:HadBoxCounts}. Total background yield in the 2b-tagged sideband is normalized to the number of observed data events in the sideband, in order to derive the scale factor  $f_{R^2,\,\mathrm{HAD}}=1.10\pm0.13$,  as described in Section~\ref{sec:ana}. The distributions of $R$ and $M_{R}$
observed in the 2b-tagged HAD box are compared to the background
prediction in Fig.~\ref{fig:signalbox}.

\begin{table}[!ht]
\begin{center}
\topcaption{Comparison of the yields in the 2b-tagged (signal region)
  samples in the HAD box.
  The uncertainties include the systematic uncertainty in the background shapes
  (Section~\ref{sec:systematics}) and statistical uncertainties. The
  uncertainty in the total yield after scaling also includes the jet
  energy scale uncertainty. The scale factor derived in the sideband
  ($f_{R^2,\,\mathrm{HAD}}=1.10\pm0.13$) is used to normalize the background yield
  in the signal-like region. The uncertainty in $f_{R^2,\,\mathrm{HAD}}$ is propagated and
  included in the quoted uncertainty in the expected background yields. }

\begin{tabular}{ccc}\hline
                                                                                   & Sideband & Signal-like\\
\hline
Multijets                                                                    & $81.0	\pm	9.5$& $34.5	\pm	6.5$\\
W/Z+jets                                                                   & $4.5	\pm	2.2$    & $11.4	\pm	3.5$\\
$\ttbar$+jets                                                          &$68.7	\pm	9.6$ & $140	\pm	11$\\
Other backgrounds                                                               & $ 0.08	\pm	0.29$&$0.16	\pm	0.48$ \\
\hline
$f_{R^2,\,\mathrm{HAD}}$                                    &  \multicolumn{2}{c|}{$1.10\pm0.13$}\\
\hline
Total background                                                   &  $170	\pm	25$& $205	\pm	28$\\
\hline\hline
Observed data                                                        &  170& 200\\

\hline
\end{tabular}

\label{tab:HadBoxCounts}
\end{center}
\end{table}

As seen in Fig.~\ref{fig:signalbox} and Table~\ref{tab:HadBoxCounts}, both the
number of observed events and the shapes of the $R$ and $M_{R}$ distributions are in
 agreement with the expected SM backgrounds. Therefore, we proceed to define two signal regions, to enhance the sensitivity for different LQ$_3$ masses. The regions are optimized to provide the lowest expected cross section limits, by varying the thresholds on $R$ and $M_{R}$.  We find that $M_{R}>400$\GeV provides the best sensitivity for all masses, and for LQ$_3$ masses below 350\GeV the optimal selection is $R^2>0.25$, while for higher masses $R^2 > 0.42$ provides best sensitivity. Because of the high value assumed for the \neutralino mass in the \sbottom search, the inclusive selection of $M_{R}>400$\GeV and $R^2>0.25$ is found to provide the optimal sensitivity in the mass range considered in this analysis.

\begin{figure}[ht!]
  \centering
  \includegraphics[width=0.6\textwidth]{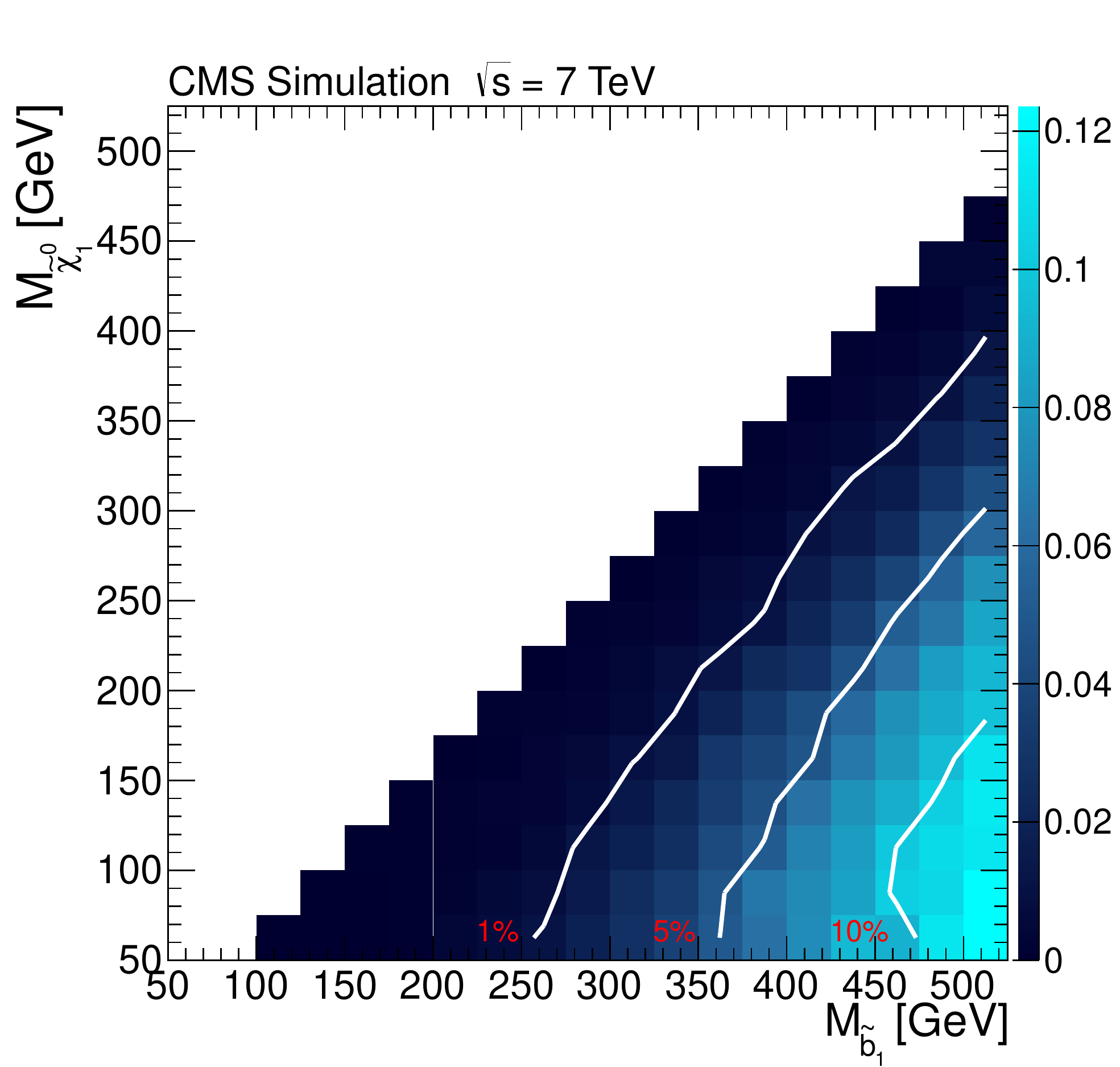}
  \caption{Signal efficiency for simulated \sbottom ~signal events with $M_{R}>400$\GeV and $R^2>0.25$. White lines show the iso-efficiency contours for 1, 5,
    and 10\% signal efficiency, respectively.}
  \label{Figure:SbottomSignalEfficiency}
\end{figure}

Table~\ref{tab:limitOptim} shows the comparison of the expected background yields in these signal boxes, and agreement of the observed event counts with the expectations is observed. Table~\ref{tab:LQ3_yields} shows the efficiency of these selections for several LQ$_3$ mass hypotheses, based on MC simulation. Efficiencies for the \sbottom signal are shown in Fig.~\ref{Figure:SbottomSignalEfficiency}. Typical efficiencies range
from a few percent up to ${\sim}12$ percent for \sbottom masses between 200 and 500\GeV and small \neutralino mass. The efficiency drops when
the mass of the \sbottom squark is close to the mass of \neutralino, since the resulting b jets are softer in these scenarios.

\begin{table}[!ht]
\begin{center}
\topcaption{Expected and observed yields in the 2b-tagged HAD box for various $R^2$ selections and a fixed $M_{R}>400$\GeV requirement. The quoted uncertainties on the expected number of events include statistical and systematic uncertainties, and the uncertainty from the $f_{R^2,\,\mathrm{HAD}}$. The left three columns show inclusive yields above the $R^2$ threshold, while the right three columns show the yields in bins of $R^2$.
  \label{tab:limitOptim}}
{
\begin{tabular}{ccccccc}\hline
$R^2$ Cut & Expected yields   & Observed yields && $R^2$ bins & Expected yields & Observed yields \\\hline
${>}0.25$	&	$205	\pm	28$	&	200   && 0.25--0.30  &	$105 \pm 25$   &  97\\
${>}0.30$	&	$100	\pm	16$	&	103   && 0.30--0.35  &         $44 \pm 11$       &  49\\
${>}0.35$	&	$56	\pm	12$	&	54     && 0.35--0.38  &         $13 \pm 9$ &  14\\
${>}0.38$	&	$43	\pm	9$	          &	40	&& 0.38--0.42  &          $18   \pm 6$       &  13\\
${>}0.42$	&	$25	\pm	7$	          &	27     && ${>}0.42$    &       $25 \pm 7$ &   27\\\hline
\end{tabular}
}
\end{center}
\end{table}

\begin{table}[!ht]
\begin{center}
\topcaption{Summary of the expected LQ$_3$ signal yields and efficiency in
  the signal region, for 4.7\fbinv of observed data, in events with $M_{R} >400$\GeV. For LQ$_3$ masses below 350\GeV $R^2>0.25$ is required, while for heavier masses we require events to pass $R^2 > 0.42$. All uncertainties are statistical only. }

{
\begin{tabular}{cccc}\hline
${M}_{\mathrm{LQ}_3}$ [\GeVns{}] & $\sigma$ (pb)    &  Efficiency (\%) & \begin{tabular}{@{}c@{}}Number \\ of expected events\end{tabular}   \\\hline
200   &  12  &  0.33  &  $185 \pm 13$ \\
250   &  3.5  &  1.1  &  $171.2 \pm 9.1$   \\
280   &  1.8  &  1.8  &  $151.4 \pm 3.6$   \\
320   &  0.82 &  3.2  &  $122.8 \pm 1.9$  \\
350   &  0.48 &  1.8   &  $39.2 \pm 1.3$ \\
450   &  0.095   &  4.3  &  $19.17 \pm 0.38$  \\
550   &  0.024   &  5.9  &  $6.59 \pm 0.12$  \\
\hline
\end{tabular}
}
\label{tab:LQ3_yields}
\end{center}
\end{table}

The statistical model for the observed number of events is a Poisson distribution with the expected value equal to the sum of the signal and expected backgrounds.  Log-normal priors for the nuisance parameters are used to model the systematic uncertainties listed in Section~\ref{sec:systematics}.

\begin{figure}[ht!]
  \begin{center}
    \includegraphics[width=0.495\textwidth]{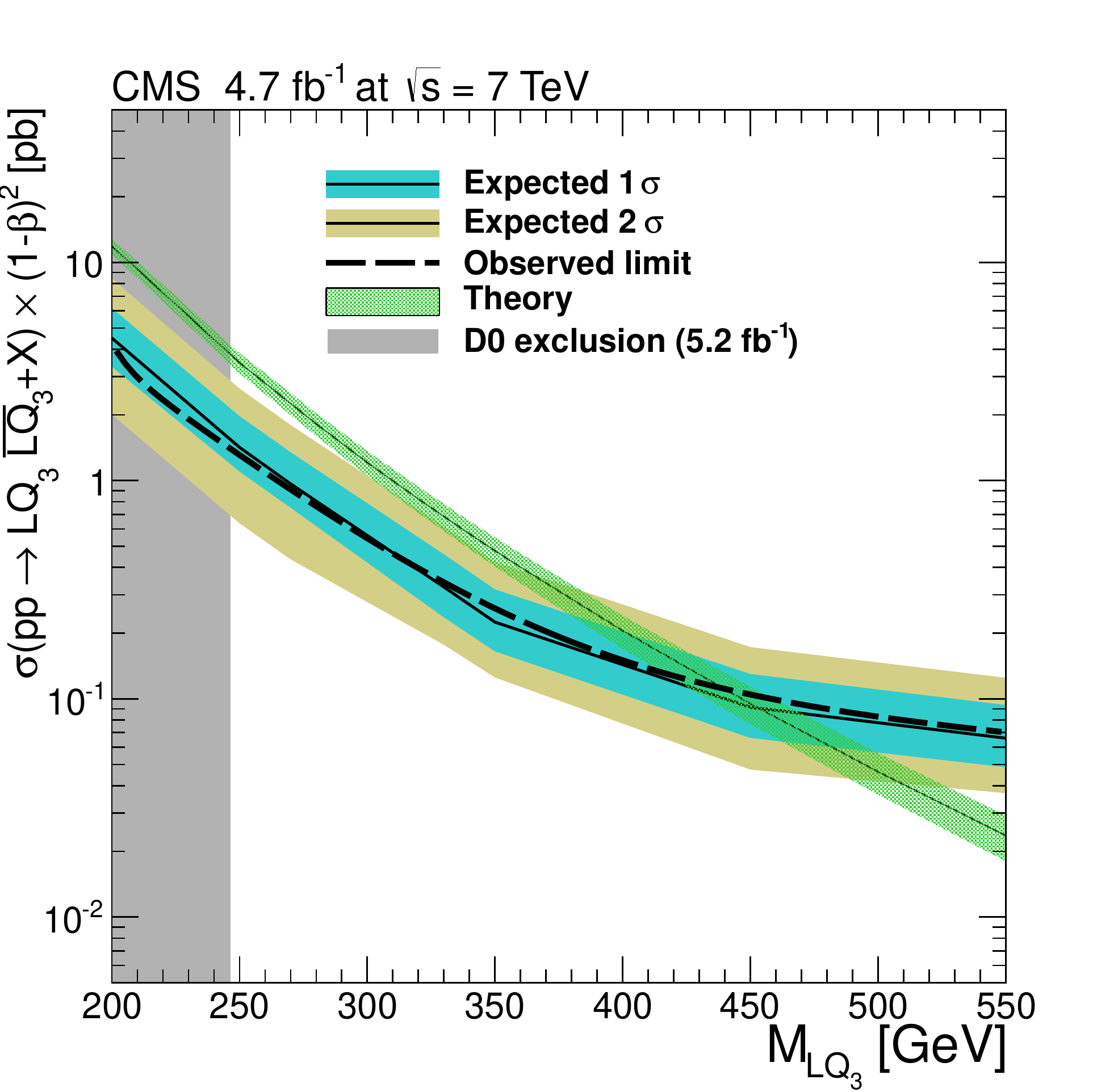}
        \includegraphics[width=0.495\textwidth]{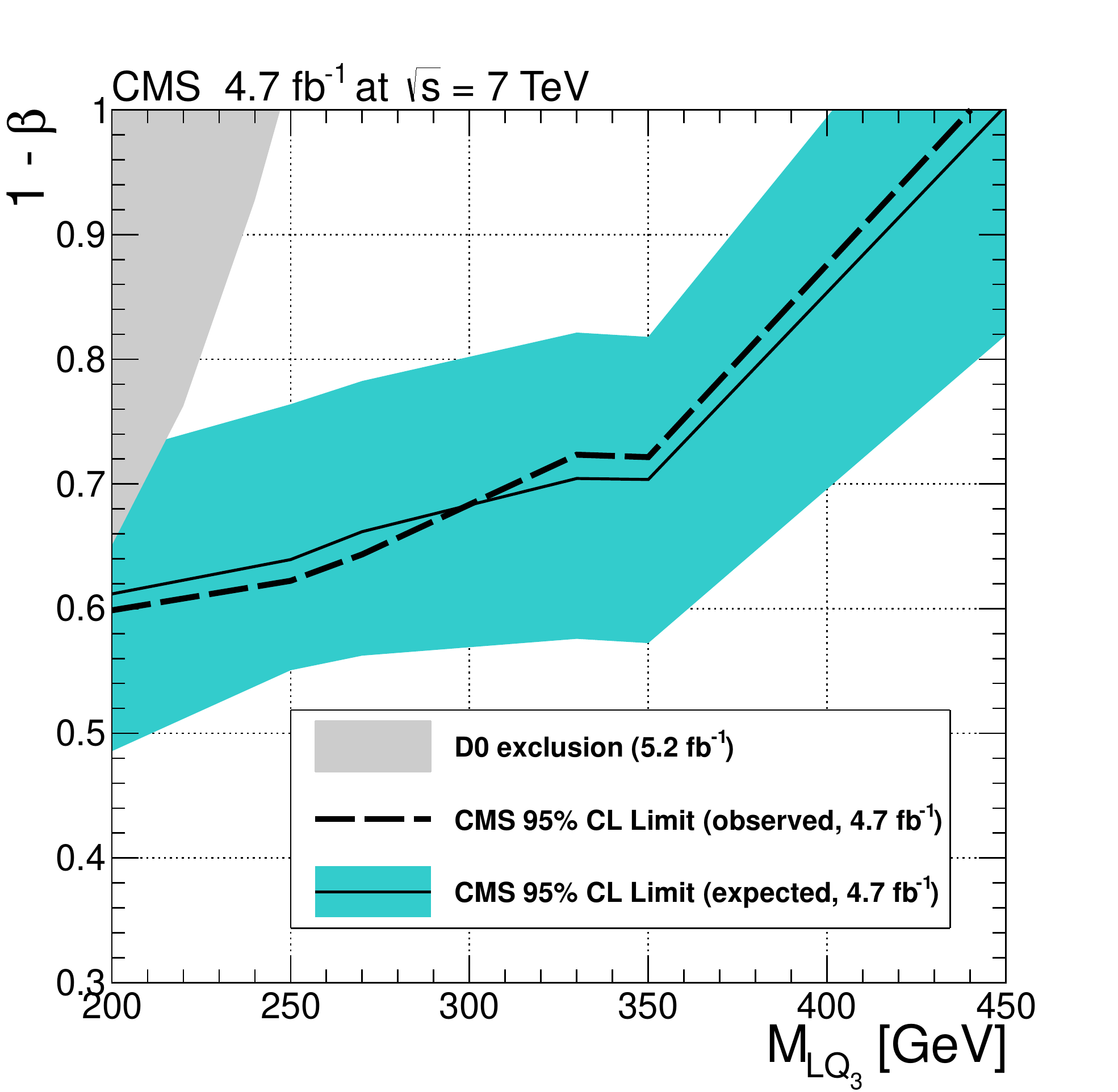}
    \caption{(Left) the expected and observed upper limit at 95\% CL on the LQ$_3$ pair production
    cross section as a function of the LQ$_3$ mass, assuming $\beta=0$. The systematic uncertainties reported in
    Section~\ref{sec:systematics} are included in the calculation. The vertical greyed region
    is excluded by the current D0 limit~\cite{Abazov201095} in the same channel. The theory curve
    and its band represent, respectively, the theoretical LQ$_3$ pair production cross section and
    the uncertainties due to the choice of parton distribution functions and renormalization/factorization
    scales~\cite{Kramer:2004df}. (Right) minimum $\beta$ for a 95\% CL exclusion of the LQ$_3$ hypothesis
    as a function of LQ$_3$ mass. The observed (expected) exclusion curve is obtained using the
    observed (expected) upper limit and the central value of the theoretical LQ$_3$ pair production
    cross section. The band around the observed exclusion curve is obtained by considering the
    observed upper limit while taking into account the uncertainties on the theoretical cross
    section. The grey region is excluded by the current D0 limits~\cite{Abazov201095} in the same channel.}
    \label{fig:limitCurves}
  \end{center}
\end{figure}

A 95\% CL upper limit is set on the potential signal cross section, as
summarized in Table~\ref{tab:limits}. The modified frequentist construction CL$_{\rm s}$~\cite{CLS, Junk:1999kv} is used for limit calculation.  These limits are interpreted in terms of limits of LQ$_3$ pair production cross section
as shown in Fig.~\ref{fig:limitCurves}. The upper limits are compared to
the NLO prediction of the LQ pair production cross
section~\cite{Kramer:2004df}, and we set a 95\% CL exclusion on LQ
masses smaller than 440\GeV (expected 470\GeV), assuming $\beta=0$.  We also present the 95\% CL limit on $\beta$ as a function of LQ$_3$ mass as shown on the right side of Fig.~\ref{fig:limitCurves}.

\begin{table}[!ht]
\begin{center}
\topcaption{Observed and expected 95\% CL upper limits on the LQ$_3$ pair-production cross section as a function of the LQ$_3$ mass.}

{
\begin{tabular}{ccccccc}\hline
${M}_{\mathrm{LQ}_3}$ [\GeVns{}]	&	-2$\sigma$	&	-1$\sigma$	&	 \begin{tabular}{@{}c@{}}Median \\ expected limit [pb]	\end{tabular}  &+1$\sigma$&+2$\sigma$&\begin{tabular}{@{}c@{}}Observed \\  limit [pb]	\end{tabular}	\\	\hline

200      &      2.0        &       3.3        &       4.5        &     6.2  &       8.4       &       4.3        \\
250 & 0.64 & 1.1 & 1.4 & 2.0 & 2.6 & 1.3 \\
270       & 0.43 & 0.75 & 0.97 & 1.4 & 1.8 & 0.90 \\
330   & 0.18 & 0.24 & 0.33 & 0.46 & 0.62 & 0.36 \\
350 & 0.13 & 0.17 & 0.23 & 0.32 & 0.42 & 0.25 \\
450   & 0.047 & 0.067 & 0.092 & 0.13 & 0.17 & 0.10 \\
550 & 0.037 & 0.049 & 0.066 & 0.094 & 0.13 & 0.073 \\

\hline
\end{tabular}
}
\label{tab:limits}
\end{center}
\end{table}

The results of the analysis are interpreted in the context of the simplified supersymmetry model spectra (SMS)~\cite{Alves:2011wf, Alwall:2008ag, Alwall:2008va}. In SMS, a limited set of hypothetical particles and decay chains are introduced to produce a given topological signature, such as the $\etmiss$ plus b jets final state considered in this analysis. We consider a SMS scenario where all supersymmetric particles   are set to have a very large mass, except for the \sbottom and \neutralino. The pairs of scalar bottom quarks produced through strong interactions are kinematically allowed to decay only into a b quark and a \neutralino.

\begin{figure}
  \begin{center}
    \includegraphics[width=0.8\textwidth]{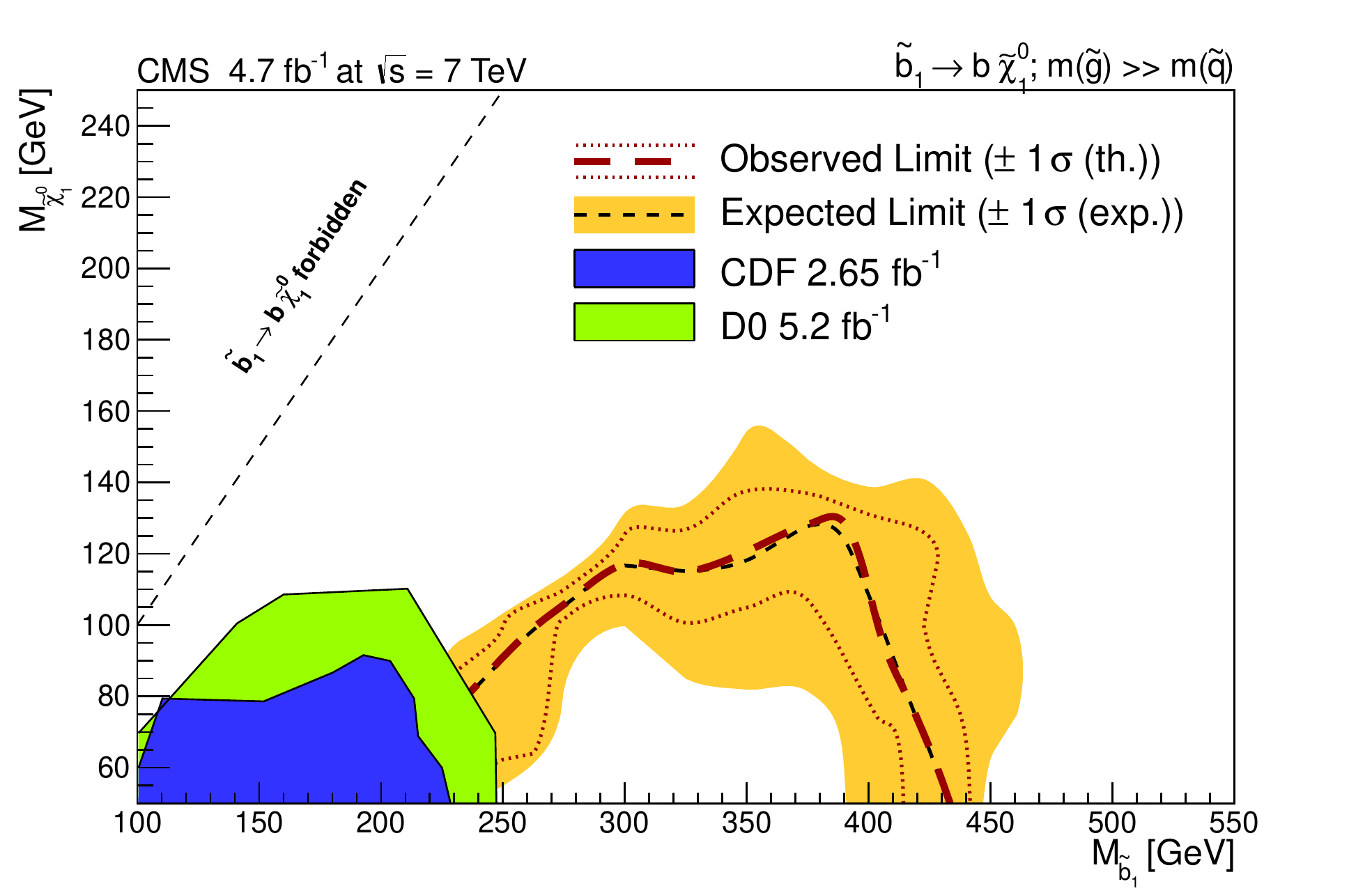}
    \caption{The expected and observed 95\% CL exclusion limits for the \sbottom pair production SMS model. The red dashed contour shows the 95\% CL exclusion limits based on the NLO+NLL cross section. The red dotted contours represent the theoretical uncertainties from the variation of parton distribution functions, and renormalization and factorization scales. The corresponding expected limits are shown with the black dashed contour. The shaded yellow contours represent the uncertainties in the SM background estimates, as reported in Section~\ref{sec:systematics}.}
    \label{Figure:ExclusionPlane}
  \end{center}
\end{figure}

The observed and expected 95\% CL upper limits in the $\tilde{\cPqb}_1-\neutralino$~mass plane are shown in Fig.~\ref{Figure:ExclusionPlane}, where the  \sbottom pair production cross section is calculated at the NLO and next-to-leading-logarithm (NLL) order~\cite{Beenakker:1996ch, Kulesza:2008jb, Kulesza:2009kq, Beenakker:2009ha, Beenakker:2011fu, Kramer:2012bx}. Since $M_{R}$ depends on the squared difference of the masses of \sbottom and \neutralino, at the \sbottom masses around 400-450\GeV and low \neutralino masses the exclusion limit is almost independent of the  \neutralino mass. The signal acceptance in the region with small mass splitting between the \sbottom and \neutralino is particularly susceptible to uncertainties associated with initial-state radiation (ISR). The impact of ISR is estimated by comparing the results of the acceptance calculation
using \PYTHIA with the ``power shower'' and with moderate ISR settings~\cite{Sjostrand:2006za}.  If the acceptance
varies by more than 25\% for a particular choice of \sbottom and \neutralino masses, then no limit is set for those mass parameters. This procedure results in reduced sensitivity in the region of $m(\sbottom)<300$\GeV and $80<m(\neutralino)<130$\GeV, and thus an inability to exclude some of the models in this parameter range.

\section{Summary}
A search has been performed for third-generation scalar leptoquarks and for scalar
bottom quarks in the all-hadronic channel with a signature of large
$\etmiss$ and b-tagged jets. This search is based on a data sample collected in pp collisions at $\sqrt{s}= 7$\TeV and corresponding to an integrated luminosity of 4.7\fbinv. The number
of observed events is in  agreement with the predictions for the SM
backgrounds.
We set an upper limit on the LQ$_3$ pair production cross section, excluding a scalar LQ$_3$
with mass below 450\GeV, assuming a 100\% branching fraction of the LQ$_3$ to b quarks and tau neutrinos.
We set 95\%  confidence level upper limits in the $\tilde{\cPqb}_1-\neutralino$ mass plane such that  for neutralino masses of 50\GeV, scalar bottom masses up to 410\GeV are excluded. These results represent the most stringent limits on LQ$_3$ masses and extend limits on \sbottom masses to much higher values than probed previously.

\section*{Acknowledgements}

We congratulate our colleagues in the CERN accelerator departments for the excellent performance of the LHC machine. We thank the technical and administrative staff at CERN and other CMS institutes, and acknowledge support from BMWF and FWF (Austria); FNRS and FWO (Belgium); CNPq, CAPES, FAPERJ, and FAPESP (Brazil); MEYS (Bulgaria); CERN; CAS, MoST, and NSFC (China); COLCIENCIAS (Colombia); MSES (Croatia); RPF (Cyprus); MoER, SF0690030s09 and ERDF (Estonia); Academy of Finland, MEC, and HIP (Finland); CEA and CNRS/IN2P3 (France); BMBF, DFG, and HGF (Germany); GSRT (Greece); OTKA and NKTH (Hungary); DAE and DST (India); IPM (Iran); SFI (Ireland); INFN (Italy); NRF and WCU (Korea); LAS (Lithuania); CINVESTAV, CONACYT, SEP, and UASLP-FAI (Mexico); MSI (New Zealand); PAEC (Pakistan); MSHE and NSC (Poland); FCT (Portugal); JINR (Armenia, Belarus, Georgia, Ukraine, Uzbekistan); MON, RosAtom, RAS and RFBR (Russia); MSTD (Serbia); SEIDI and CPAN (Spain); Swiss Funding Agencies (Switzerland); NSC (Taipei); ThEP, IPST and NECTEC (Thailand); TUBITAK and TAEK (Turkey); NASU (Ukraine); STFC (United Kingdom); DOE and NSF (USA). Individuals have received support from the Marie-Curie programme and the European Research Council (European Union); the Leventis Foundation; the A. P. Sloan Foundation; the Alexander von Humboldt Foundation; the Belgian Federal Science Policy Office; the Fonds pour la Formation \`a la Recherche dans l'Industrie et dans l'Agriculture (FRIA-Belgium); the Agentschap voor Innovatie door Wetenschap en Technologie (IWT-Belgium); the Ministry of Education, Youth and Sports (MEYS) of Czech Republic; the Council of Science and Industrial Research, India; the Compagnia di San Paolo (Torino); and the HOMING PLUS programme of Foundation for Polish Science, cofinanced from European Union, Regional Development Fund.

\bibliography{auto_generated}

\providecommand{\href}[2]{#2}\begingroup\raggedright\begin{thebibliography}{10}%
\makeatletter
\providecommand{\hrefCMSnoop }[0]{\@secondoftwo}%
\makeatother
\providecommand{\doi}{\texttt{doi:}\begingroup \urlstyle{tt}\Url}

\bibitem{PatiSalam}
\hrefCMSnoop {} {J.~C. Pati and A.~Salam, ``Lepton number as the fourth
  `color''',} \textit{ Phys. Rev. D} \textbf{ 10} (1974) 275,
  \href{http://dx.doi.org/10.1103/PhysRevD.10.275}{\doi{10.1103/PhysRevD.10.275}}.

\bibitem{Schrempp1985101}
\hrefCMSnoop {} {B.~Schrempp and F.~Schrempp, ``Light leptoquarks'',} \textit{
  Phys. Lett. B} \textbf{ 153} (1985) 101,
  \href{http://dx.doi.org/10.1016/0370-2693(85)91450-9}{\doi{10.1016/0370-2693(85)91450-9}}.

\bibitem{Gripaios:2010hv}
B.~Gripaios\hrefCMSnoop {} { {et~al.}, ``{Searching for third-generation
  composite leptoquarks at the LHC}'',} \textit{ J. High Energy Phys.} \textbf{
  01} (2011) 156,
  \href{http://dx.doi.org/10.1007/JHEP01(2011)156}{\doi{10.1007/JHEP01(2011)156}},
\href{http://www.arXiv.org/abs/1010.3962}{\texttt{ arXiv:1010.3962}}.

\bibitem{Dimopoulos1979237}
\hrefCMSnoop {} {S.~Dimopoulos and L.~Susskind, ``Mass without scalars'',}
  \textit{ Nucl. Phys. B} \textbf{ 155} (1979) 237,
  \href{http://dx.doi.org/10.1016/0550-3213(79)90364-X}{\doi{10.1016/0550-3213(79)90364-X}}.

\bibitem{Dimopoulos198069}
\hrefCMSnoop {} {S.~Dimopoulos, ``Technicoloured signatures'',} \textit{ Nucl.
  Phys. B} \textbf{ 168} (1980) 69,
  \href{http://dx.doi.org/10.1016/0550-3213(80)90277-1}{\doi{10.1016/0550-3213(80)90277-1}}.

\bibitem{Farhi1981277}
\hrefCMSnoop {} {E.~Farhi and L.~Susskind, ``Technicolour'',} \textit{ Phys.
  Rep.} \textbf{ 74} (1981) 277,
  \href{http://dx.doi.org/10.1016/0370-1573(81)90173-3}{\doi{10.1016/0370-1573(81)90173-3}}.

\bibitem{Hewett1989193}
\hrefCMSnoop {} {J.~L. Hewett and T.~G. Rizzo, ``Low-energy phenomenology of
  superstring-inspired E$_6$ models'',} \textit{ Phys. Rep.} \textbf{ 183}
  (1989) 193,
  \href{http://dx.doi.org/10.1016/0370-1573(89)90071-9}{\doi{10.1016/0370-1573(89)90071-9}}.

\bibitem{DavidsonBaileyCampbell}
\hrefCMSnoop {} {S.~Davidson, D.~C. Bailey, and B.~A. Campbell, ``{Model
  independent constraints on leptoquarks from rare processes}'',} \textit{ Z.
  Phys. C} \textbf{ 61} (1994) 613,
  \href{http://dx.doi.org/10.1007/BF01552629}{\doi{10.1007/BF01552629}},
\href{http://www.arXiv.org/abs/hep-ph/9309310}{\texttt{ arXiv:hep-ph/9309310}}.

\bibitem{Farrar1978575}
\hrefCMSnoop {} {G.~R. Farrar and P.~Fayet, ``Phenomenology of the production,
  decay, and detection of new hadronic states associated with supersymmetry'',}
  \textit{ Phys. Lett. B} \textbf{ 76} (1978) 575,
  \href{http://dx.doi.org/10.1016/0370-2693(78)90858-4}{\doi{10.1016/0370-2693(78)90858-4}}.

\bibitem{Feng:2010gw}
\hrefCMSnoop {} {J.~L. Feng, ``{Dark matter candidates from particle physics
  and methods of detection}'',} \textit{ Annu. Rev. Astron. Astr.} \textbf{ 48}
  (2010) 495,
\href{http://dx.doi.org/10.1146/annurev-astro-082708-101659}{\doi{10.1146/annurev-astro-082708-101659}}.

\bibitem{Dimopoulos:1995mi}
\hrefCMSnoop {} {S.~Dimopoulos and G.~F. Giudice, ``{Naturalness constraints in
  supersymmetric theories with non-universal soft terms}'',} \textit{ Phys.
  Lett. B} \textbf{ 357} (1995) 573,
  \href{http://dx.doi.org/10.1016/0370-2693(95)00961-J}{\doi{10.1016/0370-2693(95)00961-J}}.

\bibitem{Abazov201095}
\hrefCMSnoop {} {{ D0} Collaboration, ``Search for scalar bottom quarks and
  third-generation leptoquarks in $p\bar{p}$ collisions at
  $\sqrt{s}=1.96\text{\,}\text{\,}\mathrm{TeV}$'',} \textit{ Phys. Lett. B}
  \textbf{ 693} (2010) 95,
  \href{http://dx.doi.org/10.1016/j.physletb.2010.08.028}{\doi{10.1016/j.physletb.2010.08.028}}.

\bibitem{PhysRevLett.105.081802}
\hrefCMSnoop {} {{ CDF} Collaboration, ``Search for the production of scalar
  bottom quarks in $p\overline{p}$ collisions at
  $\sqrt{s}=1.96\text{\,}\text{\,}\mathrm{TeV}$'',} \textit{ Phys. Rev. Lett.}
  \textbf{ 105} (2010) 081802,
  \href{http://dx.doi.org/10.1103/PhysRevLett.105.081802}{\doi{10.1103/PhysRevLett.105.081802}}.

\bibitem{CMS-PAS-EXO-12-002}
\hrefCMSnoop {} {{CMS Collaboration}, ``Search for pair production of third
  generation leptoquarks and stops that decay to a tau and a b quark'',}
  (2012). \href{http://www.arXiv.org/abs/1210.5629}{\texttt{ arXiv:1210.5629}}.

\bibitem{PhysRevLett.108.181802}
\hrefCMSnoop {} {{ ATLAS} Collaboration, ``Search for scalar bottom quark pair
  production with the ATLAS detector in pp collisions at
  $\sqrt{s}=7\text{\,}\text{\,}\mathrm{TeV}$'',} \textit{ Phys. Rev. Lett.}
  \textbf{ 108} (2012) 181802,
  \href{http://dx.doi.org/10.1103/PhysRevLett.108.181802}{\doi{10.1103/PhysRevLett.108.181802}}.

\bibitem{PTDR1}
\hrefCMSnoop {} {{CMS Collaboration}, ``The CMS experiment at the CERN LHC'',}
  \textit{ J. Instrum.} \textbf{ 3} (2008) S08004,
  \href{http://dx.doi.org/10.1088/1748-0221/3/08/S08004}{\doi{10.1088/1748-0221/3/08/S08004}}.

\bibitem{rogan}
\hrefCMSnoop {} {C.~Rogan, ``Kinematics for new dynamics at the LHC'',} (2010).
  \href{http://www.arXiv.org/abs/1006.2727}{\texttt{ arXiv:1006.2727}}.

\bibitem{SUS-10-009}
\hrefCMSnoop {} {{ CMS} Collaboration, ``Inclusive search for squarks and
  gluinos in {$\Pp\Pp$} collisions at {$\sqrt{s}=7\TeV$}'',} \textit{ Phys.
  Rev. D} \textbf{ 85} (2011) 012004,
  \href{http://dx.doi.org/10.1103/PhysRevD.85.012004}{\doi{10.1103/PhysRevD.85.012004}}.

\bibitem{Sjostrand:2006za}
\hrefCMSnoop {} {T.~Sj{\"{o}}strand, S.~Mrenna, and P.~Z. Skands, ``{\PYTHIA
  6.4 physics and manual}'',} \textit{ J. High Energy Phys.} \textbf{ 05}
  (2006) 026,
\href{http://dx.doi.org/10.1088/1126-6708/2006/05/026}{\doi{10.1088/1126-6708/2006/05/026}}.

\bibitem{madgraph}
J.~Alwall\hrefCMSnoop {} { {et~al.}, ``{\MADGRAPH 5 : going beyond}'',}
  \textit{ J. High Energy Phys.} \textbf{ 06} (2011) 128,
\href{http://dx.doi.org/10.1007/JHEP06(2011)128}{\doi{10.1007/JHEP06(2011)128}}.

\bibitem{G4}
\hrefCMSnoop {} {{ GEANT4} Collaboration, ``{\GEANT4---a simulation
  toolkit}'',} \textit{ Nucl. Instrum. Meth. A} \textbf{ 506} (2003) 250,
\href{http://dx.doi.org/10.1016/S0168-9002(03)01368-8}{\doi{10.1016/S0168-9002(03)01368-8}}.

\bibitem{Chatrchyan:2011id}
\hrefCMSnoop {} {{ CMS} Collaboration, ``{Measurement of the underlying event
  activity at the LHC with $\sqrt{s}= 7$ TeV and comparison with $\sqrt{s} =
  0.9$ TeV}'',} \textit{ J. High Energy Phys.} \textbf{ 09} (2011) 109,
\href{http://dx.doi.org/10.1007/JHEP09(2011)109}{\doi{10.1007/JHEP09(2011)109}}.

\bibitem{Field:2008zz}
\href {http://th-www.if.uj.edu.pl/acta/vol39/abs/v39p2611.htm} {R.~Field,
  ``{Physics at the Tevatron}'',} \textit{ Acta Phys. Pol. B} \textbf{
  \href{http://th-www.if.uj.edu.pl/acta/vol39/abs/v39p2611.htm}{39}} (2008)
2611.

\bibitem{Field:2010su}
\href {http://www-library.desy.de/preparch/desy/proc/proc09-06.pdf} {R.~Field,
  ``{Studying the underlying event at CDF and the LHC}'',} in \textit{
  Proceedings of the First International Workshop on Multiple Partonic
  Interactions at the LHC MPI'08, October 27-31, 2008}, P.~Bartalini and
  L.~Fan{\'o}, eds., p.~12.
\newblock October, 2009.
\newblock \href{http://www.arXiv.org/abs/1003.4220}{\texttt{ arXiv:1003.4220}}.

\bibitem{CMS-DP-2010-039}
\href {http://cdsweb.cern.ch/record/1309890} {{ CMS} Collaboration,
  ``Comparison of the fast simulation of CMS with the first LHC data'',} CMS
  Detector Performance Summary CMS-DP-2010-039, (2010).

\bibitem{TRK-10-005}
\href {http://cdsweb.cern.ch/record/1279383} {{ CMS} Collaboration, ``Tracking
  and Primary Vertex Results in First 7 {TeV} Collisions'',} CMS Physics
  Analysis Summary CMS-PAS-TRK-10-005, (2010).

\bibitem{METJINST}
\hrefCMSnoop {} {{ CMS} Collaboration, ``Missing transverse energy performance
  of the {CMS} detector'',} \textit{ J. Instrum.} \textbf{ 6} (2011) P09001,
  \href{http://dx.doi.org/10.1088/1748-0221/6/09/P09001}{\doi{10.1088/1748-0221/6/09/P09001}}.

\bibitem{antikt}
\hrefCMSnoop {} {M.~Cacciari, G.~P. Salam, and G.~Soyez, ``{The anti-$k_t$ jet
  clustering algorithm}'',} \textit{ J. High Energy Phys.} \textbf{ 04} (2008)
  063,
\href{http://dx.doi.org/10.1088/1126-6708/2008/04/063}{\doi{10.1088/1126-6708/2008/04/063}}.

\bibitem{cms:jes}
\hrefCMSnoop {} {{ CMS} Collaboration, ``Determination of jet energy
  calibration and transverse momentum resolution in {CMS}'',} \textit{ J.
  Instrum.} \textbf{ 6} (2011) P11002,
  \href{http://dx.doi.org/10.1088/1748-0221/6/11/P11002}{\doi{10.1088/1748-0221/6/11/P11002}}.

\bibitem{CMS-PAS-PFT-10-002}
\href {http://cdsweb.cern.ch/record/1279341} {{ CMS} Collaboration,
  ``Commissioning of the Particle-Flow Reconstruction in Minimum-Bias and Jet
  Events from {\Pp\Pp} Collisions at 7 {TeV}'',} CMS Physics Analysis Summary
  CMS-PAS-PFT-10-002, (2010).

\bibitem{CMSWZxsections}
\hrefCMSnoop {} {{ CMS} Collaboration, ``{Measurements of inclusive W and Z
  cross sections in pp collisions at $\sqrt{s} =7$~TeV}'',} \textit{ J. High
  Energy Phys.} \textbf{ 01} (2011) 080,
\href{http://dx.doi.org/10.1007/JHEP01(2011)080}{\doi{10.1007/JHEP01(2011)080}}.

\bibitem{Cacciari:2007fd}
\hrefCMSnoop {} {M.~Cacciari and G.~P. Salam, ``{Pileup subtraction using jet
  areas}'',} \textit{ Phys. Lett. B} \textbf{ 659} (2008) 119,
  \href{http://dx.doi.org/10.1016/j.physletb.2007.09.077}{\doi{10.1016/j.physletb.2007.09.077}}.

\bibitem{CMS-PAS-BTV-11-004}
\href {http://cdsweb.cern.ch/record/1427247} {{ CMS} Collaboration, ``b-Jet
  Identification in the {CMS} Experiment'',} CMS Physics Analysis Summary
  CMS-PAS-BTV-11-004, (2011).

\bibitem{Chatrchyan:2011nb}
\hrefCMSnoop {} {{ CMS} Collaboration, ``{Measurement of the $t\bar{t}$
  production cross section and the top quark mass in the dilepton channel in pp
  collisions at $\sqrt{s} =7$~TeV}'',} \textit{ J. High Energy Phys.} \textbf{
  07} (2011) 049,
  \href{http://dx.doi.org/10.1007/JHEP07(2011)049}{\doi{10.1007/JHEP07(2011)049}}.

\bibitem{verkerke-2003}
\hrefCMSnoop {} {W.~{Verkerke} and D.~{Kirkby}, ``{The {\sc RooFit} toolkit for
  data modeling}'',} (2003).
  \href{http://www.arXiv.org/abs/physics/0306116}{\texttt{
  arXiv:physics/0306116}}.

\bibitem{CMS-PAS-EWK-11-017}
\href {http://cdsweb.cern.ch/record/1431015} {{ CMS} Collaboration, ``Study of
  the dijet invariant mass distribution in {W}$\to\ell\nu$ plus jets events
  produced in pp collisions at $\sqrt{s} = 7$~{TeV}'',} CMS Physics Analysis
  Summary CMS-PAS-EWK-11-017, (2011).

\bibitem{Chatrchyan:2012vr}
\hrefCMSnoop {} {{ CMS} Collaboration, ``{Measurement of the Z/gamma*+b-jet
  cross section in pp collisions at 7 TeV}'',} \textit{ J. High Energy Phys.}
  \textbf{ 06} (2012) 126,
\href{http://dx.doi.org/10.1007/JHEP06(2012)126}{\doi{10.1007/JHEP06(2012)126}}.

\bibitem{Botje:2011sn}
M.~Botje\hrefCMSnoop {} { {et~al.}, ``{The PDF4LHC working group interim
  recommendations}'',} (2011).
\href{http://www.arXiv.org/abs/1101.0538}{\texttt{ arXiv:1101.0538}}.

\bibitem{Martin:2009iq}
A.~D. Martin\hrefCMSnoop {} { {et~al.}, ``{Parton distributions for the
  LHC}'',} \textit{ Eur. Phys. J. C} \textbf{ 63} (2009) 189,
\href{http://dx.doi.org/10.1140/epjc/s10052-009-1072-5}{\doi{10.1140/epjc/s10052-009-1072-5}}.

\bibitem{Nadolsky:2008zw}
P.~M. Nadolsky\hrefCMSnoop {} { {et~al.}, ``{Implications of CTEQ global
  analysis for collider observables}'',} \textit{ Phys. Rev. D} \textbf{ 78}
  (2008) 013004,
\href{http://dx.doi.org/10.1103/PhysRevD.78.013004}{\doi{10.1103/PhysRevD.78.013004}}.

\bibitem{Ball:2010de}
R.~D. Ball\hrefCMSnoop {} { {et~al.}, ``{A first unbiased global NLO
  determination of parton distributions and their uncertainties}'',} \textit{
  Nucl. Phys. B} \textbf{ 838} (2010) 136,
\href{http://dx.doi.org/10.1016/j.nuclphysb.2010.05.008}{\doi{10.1016/j.nuclphysb.2010.05.008}}.

\bibitem{PDFLQ3}
J.~Pumplin\hrefCMSnoop {} { {et~al.}, ``{New generation of parton distributions
  with uncertainties from global QCD analysis}'',} \textit{ J. High Energy
  Phys.} \textbf{ 07} (2002) 012,
  \href{http://dx.doi.org/10.1088/1126-6708/2002/07/012}{\doi{10.1088/1126-6708/2002/07/012}},
\href{http://www.arXiv.org/abs/hep-ph/0201195}{\texttt{ arXiv:hep-ph/0201195}}.

\bibitem{Martin:2002aw}
A.~D. Martin\hrefCMSnoop {} { {et~al.}, ``Uncertainties of predictions from
  parton distributions I: Experimental errors'',} \textit{ Eur. Phys. J. C}
  \textbf{ 28} (2003) 455,
\href{http://dx.doi.org/10.1140/epjc/s2003-01196-2}{\doi{10.1140/epjc/s2003-01196-2}}.

\bibitem{Kramer:2004df}
M.~Kr{\"a}mer\hrefCMSnoop {} { {et~al.}, ``{Pair production of scalar
  leptoquarks at the LHC}'',} \textit{ Phys. Rev. D} \textbf{ 71} (2005)
  057503,
\href{http://dx.doi.org/10.1103/PhysRevD.71.057503}{\doi{10.1103/PhysRevD.71.057503}}.

\bibitem{CMS-PAS-SMP-12-008}
\href {http://cdsweb.cern.ch/record/1434360} {{ CMS} Collaboration, ``Absolute
  Calibration of the Luminosity Measurement at {CMS}: {W}inter 2012 Update'',}
  CMS Physics Analysis Summary CMS-PAS-SMP-12-008, (2012).

\bibitem{CLS}
\hrefCMSnoop {} {A.~L. Read, ``Presentation of search results: the $CL_s$
  technique'',} \textit{ J. Phys. G} \textbf{ 28} (2002) 2693,
  \href{http://dx.doi.org/10.1088/0954-3899/28/10/313}{\doi{10.1088/0954-3899/28/10/313}}.

\bibitem{Junk:1999kv}
\hrefCMSnoop {} {T.~Junk, ``{Confidence level computation for combining
  searches with small statistics}'',} \textit{ Nucl. Instrum. Meth. A} \textbf{
  434} (1999) 435,
\href{http://dx.doi.org/10.1016/S0168-9002(99)00498-2}{\doi{10.1016/S0168-9002(99)00498-2}}.

\bibitem{Alves:2011wf}
D.~Alves\hrefCMSnoop {} { {et~al.}, ``{Simplified models for LHC new physics
  searches}'',} (2011).
\href{http://www.arXiv.org/abs/1105.2838}{\texttt{ arXiv:1105.2838}}.

\bibitem{Alwall:2008ag}
\hrefCMSnoop {} {J.~Alwall, P.~Schuster, and N.~Toro, ``{Simplified models for
  a first characterization of new physics at the LHC}'',} \textit{ Phys. Rev.
  D} \textbf{ 79} (2009) 075020,
\href{http://dx.doi.org/10.1103/PhysRevD.79.075020}{\doi{10.1103/PhysRevD.79.075020}}.

\bibitem{Alwall:2008va}
J.~Alwall\hrefCMSnoop {} { {et~al.}, ``{Model-independent jets plus missing
  energy searches}'',} \textit{ Phys. Rev. D} \textbf{ 79} (2009) 015005,
\href{http://dx.doi.org/10.1103/PhysRevD.79.015005}{\doi{10.1103/PhysRevD.79.015005}}.

\bibitem{Beenakker:1996ch}
W.~Beenakker\hrefCMSnoop {} { {et~al.}, ``{Squark and gluino production at
  hadron colliders}'',} \textit{ Nucl. Phys. B} \textbf{ 492} (1997) 51,
  \href{http://dx.doi.org/10.1016/S0550-3213(97)00084-9}{\doi{10.1016/S0550-3213(97)00084-9}}.

\bibitem{Kulesza:2008jb}
\hrefCMSnoop {} {A.~Kulesza and L.~Motyka, ``{Threshold resummation for
  squark-antisquark and gluino-pair production at the LHC}'',} \textit{ Phys.
  Rev. Lett.} \textbf{ 102} (2009) 111802,
  \href{http://dx.doi.org/10.1103/PhysRevLett.102.111802}{\doi{10.1103/PhysRevLett.102.111802}}.

\bibitem{Kulesza:2009kq}
\hrefCMSnoop {} {A.~Kulesza and L.~Motyka, ``{Soft gluon resummation for the
  production of gluino-gluino and squark-antisquark pairs at the LHC}'',}
  \textit{ Phys. Rev. D} \textbf{ 80} (2009) 095004,
  \href{http://dx.doi.org/10.1103/PhysRevD.80.095004}{\doi{10.1103/PhysRevD.80.095004}}.

\bibitem{Beenakker:2009ha}
W.~Beenakker\hrefCMSnoop {} { {et~al.}, ``{Soft-gluon resummation for squark
  and gluino hadroproduction}'',} \textit{ J. High Energy Phys.} \textbf{ 0912}
  (2009) 041,
  \href{http://dx.doi.org/10.1088/1126-6708/2009/12/041}{\doi{10.1088/1126-6708/2009/12/041}}.

\bibitem{Beenakker:2011fu}
W.~Beenakker\hrefCMSnoop {} { {et~al.}, ``{Squark and gluino
  hadroproduction}'',} \textit{ Int. J. Mod. Phys. A} \textbf{ 26} (2011) 2637,
  \href{http://dx.doi.org/10.1142/S0217751X11053560}{\doi{10.1142/S0217751X11053560}}.

\bibitem{Kramer:2012bx}
M.~Kr{\"a}mer\hrefCMSnoop {} { {et~al.}, ``{Supersymmetry production cross
  sections in pp collisions at $\sqrt{s} =7$~TeV}'',} (2012).
\href{http://www.arXiv.org/abs/1206.2892}{\texttt{ arXiv:1206.2892}}.

\end{thebibliography}\endgroup

\cleardoublepage \appendix\section{The CMS Collaboration \label{app:collab}}\begin{sloppypar}\hyphenpenalty=5000\widowpenalty=500\clubpenalty=5000\textbf{Yerevan Physics Institute,  Yerevan,  Armenia}\\*[0pt]
S.~Chatrchyan, V.~Khachatryan, A.M.~Sirunyan, A.~Tumasyan
\vskip\cmsinstskip
\textbf{Institut f\"{u}r Hochenergiephysik der OeAW,  Wien,  Austria}\\*[0pt]
W.~Adam, E.~Aguilo, T.~Bergauer, M.~Dragicevic, J.~Er\"{o}, C.~Fabjan\cmsAuthorMark{1}, M.~Friedl, R.~Fr\"{u}hwirth\cmsAuthorMark{1}, V.M.~Ghete, N.~H\"{o}rmann, J.~Hrubec, M.~Jeitler\cmsAuthorMark{1}, W.~Kiesenhofer, V.~Kn\"{u}nz, M.~Krammer\cmsAuthorMark{1}, I.~Kr\"{a}tschmer, D.~Liko, I.~Mikulec, M.~Pernicka$^{\textrm{\dag}}$, D.~Rabady\cmsAuthorMark{2}, B.~Rahbaran, C.~Rohringer, H.~Rohringer, R.~Sch\"{o}fbeck, J.~Strauss, A.~Taurok, W.~Waltenberger, C.-E.~Wulz\cmsAuthorMark{1}
% ...
\vskip\cmsinstskip
\dag:~Deceased\\
1:~~Also at Vienna University of Technology, Vienna, Austria\\
2:~~Also at CERN, European Organization for Nuclear Research, Geneva, Switzerland\\
% ...

\end{sloppypar}
\end{document}